\documentclass{elsarticle}
\usepackage{graphicx,color}
\usepackage{amsmath}
\usepackage{amssymb}

\newcommand \beq{\begin{eqnarray}}
\newcommand \eeq{\end{eqnarray}} 
\def\x{{\boldsymbol x}}

\def\q{{\boldsymbol q}}
\def\p{{\boldsymbol p}}

\def\nab{{\boldsymbol \nabla}}

\newcommand{\rmd}{{\rm d}}
\newcommand{\rme}{{\rm e}}

\newcommand{\del}{\partial}
\newcommand{\nn}{\nonumber\\ }

\begin{document}
\begin{frontmatter}
\title{The thermalization of soft modes in non-expanding isotropic quark gluon plasmas  }
\author[cea]{Jean-Paul Blaizot}
\author[ind,rbrc]{Jinfeng Liao}
\author[wu]{Yacine Mehtar-Tani}

\address[cea]{Institut de  Physique Th\'eorique, CNRS/UMR 3681, CEA Saclay,  F-91191 Gif-sur-Yvette, France}
\address[ind]{Physics Department and Center for Exploration of Energy and Matter,
Indiana University, 2401 N Milo B. Sampson Lane, Bloomington, IN 47408, USA}
\address[rbrc]{RIKEN BNL Research Center, Bldg. 510A, Brookhaven National Laboratory,
   Upton, NY 11973, USA}
\address[wu]{Institute for Nuclear Theory, University of Washington, 
Seattle, WA 98195-1550, USA}

\begin{abstract}
We discuss the role of elastic and inelastic collisions  and their interplay in the thermalization of the quark-gluon plasma. We consider a simplified situation of  a static plasma, spatially uniform and isotropic in momentum space. We focus on the small momentum region, which equilibrates first, and on a short time scale. We obtain a simple kinetic equation that allows for an analytic description of the most important regimes. The present analysis suggests that the formation of a Bose condensate, expected when only elastic collisions are present, is strongly hindered by the inelastic, radiative, processes. 
\end{abstract}

\begin{keyword}
Perturbative QCD, Heavy Ion Collisions, Thermalization, Bose-Einstein Condensation\\
\end{keyword}

\end{frontmatter}

 \begin{flushright}
INT-PUB-16-030
 \end{flushright}
\section{Introduction}

Understanding how the  gluons that are freed in the early stage of a heavy ion collision locally equilibrate and exhibit  collective fluid behavior observed in experiments  remains an important and challenging problem, with many interesting facets and open issues (see Ref.~\cite{Berges:2012ks,Huang:2014iwa,Fukushima:2016xgg} for reviews). Within weak coupling approaches, to which the present discussion is limited, much of the relevant physical processes have been identified. These involve various plasma instabilities and their potential role in isotropizing the momentum distribution, the elastic and inelastic scatterings, and among the latter the importance of soft radiation \cite{Baier:2000sb,Xu:2004mz},  the role of the longitudinal expansion, etc. There exist detailed \cite{Baier:2000sb}, and very detailed \cite{Kurkela:2011ti},  parametric analysis at weak coupling  of these various physical processes. Extended numerical  simulations using statistical classical field theory point out the existence of regimes dominated by non thermal fixed points, with characteristic scaling behavior (see  \cite{Berges:2013fga} for a representative example). Additional insight is provided by kinetic theory, which focusses on the direct interactions between  modes or quasiparticles \cite{Mueller:1999pi,Xu:2004mz,Scardina:2014gxa}. From some of these works, evidence is emerging that thermalization can indeed be achieved among weakly coupled gluons on reasonably short time scales (see e.g. \cite{Gelis:2013rba,Kurkela:2014tea}).

However  a fully coherent  picture is still lacking, and  many issues remain to be clarified. Furthermore all approaches have limitations. For instance, numerical simulations have difficulties to handle very long wavelength modes, and they do not always offer the physical insight that one is looking for.  As for the parametric estimates, they are blind to the possible presence of large numerical factors (and there are such factors) that can obscure the picture; one may be led for instance to expect the existence of regimes that the full solution does not reveal, because in practice, scales are not as well separated as the parametric analysis would suggest. 
Given the overall complexity of the problem, 
we feel therefore that there is a need  for  developing an understanding based on simple (differential) equations that we can control (semi) analytically. Our goal is certainly not to resolve all the pending issues, but to identify robust, generic qualitative  behaviors and understand the basic mechanisms and the factors that control the important  time scales. 
This requires simplifications, in the choice of the system to be studied, and in the choice of an appropriate theoretical framework that may allow for analytical insight.  The simplifications concerning the system to be studied are standard, and they can be alleviated with moderate effort: thus, in this paper, we shall study a uniform, non expanding system, which furthermore is isotropic in momentum space.   Moreover  we shall focus on small momentum modes. Relaxing these assumptions is possible, as just said, and will be the subject of future publications. As for the choice of the framework, we choose to  work within kinetic theory, to which we now turn. 

We shall use a kinetic equation of the generic form 
\beq\label{transport0}
\frac{\del f(t,\p)}{\del t}=C_{\rm el}[f]+C_{\rm inel}[f],
\eeq
where the collision integral is conveniently split into two contributions: one corresponding to elastic, number conserving, scattering; the other referring to inelastic, number changing, processes. Such an equation has been used, implicitly or explicitly in many discussions of the thermalization of the quark-gluon plasma (see e.g. \cite{Baier:2000sb,Arnold:2002zm,Mueller:2006up}).

The present work builds on our previous works on the subject \cite{Blaizot:2013lga} where only elastic scatterings were taken into account, i.e., $C_{\rm inel}[f]$ was set to zero. Because the dominant scattering among gluons occur at  small angle, we can reduce the Boltzmann equation to a  (much simpler)  Fokker-Planck equation, in which the collision  integral is written as the divergence of a current (in momentum space). The small scattering angle approximation has been checked against numerical  solutions of the Boltzmann equation, and no major qualitative differences could be observed \cite{Scardina:2014gxa,Xu:2014ega}.  The Fokker-Planck equation  provides a simple visualization of the physics in terms of competing currents of particles in momentum space.  A particular striking consequence of this analysis is the prediction that, for typical initial conditions,  gluons could undergo Bose-Einstein condensation (BEC), as first suggested in  \cite{Blaizot:2011xf}. Of course inelastic scattering will eventually prevent a true condensate to be present in the equilibrium state, but a priori this does not preclude the formation of a transient condensate.  Note that the existence of such a condensate in non-Abelian gauge theories raises a number of conceptual issues, some of which are addressed in \cite{Kurkela:2012hp}. 

In general, inelastic processes are expected to be suppressed in the weak coupling limit, compared to elastic processes. However, this is not the case in the case of non-Abelian plasmas,  where they turn out to be parametrically as important as the elastic ones. Aside from being of the same order of magnitude, these processes involve nearly collinear emission of soft gluons. As we shall see, it is the singular behavior of these soft emissions, rather than the mere order of magnitude of the rates, that affects most strongly the long wavelength properties of the system.  In fact, the collinear splittings involved here are quite similar to  those involved in the cascades of gluons that give rise to  jets. And we shall rely on this analogy to exploit approximations that we have used in previous work on jet physics in order to obtain a simple expression for the inelastic collision term \cite{Blaizot:2015lma}.  
The resulting  kinetic equation is a simplified version of that used in the ``bottom-up'' scenario \cite{Baier:2000sb}. A more elaborate version of this kinetic equation was solved numerically recently \cite{York:2014wja,Kurkela:2014tea}.

One may question the validity of kinetic theory, in particular when applied to the region of soft momentum modes. The conditions of validity  of kinetic theory within weak coupling approaches have been thoroughly reviewed in \cite{Arnold:2002zm}. Typically, one expects a kinetic description to be valid when the occupation number are not too large, so that non-linear effects are weak, i.e.  $f(p) \ll 1/g^2$ with $g$ the gauge coupling. When the occupation becomes too large, it is usually better to turn to a classical field description. Furthermore, kinetic theory presupposes the existence of well defined modes, or quasiparticles,  which may not be the case for too small momenta. For this reason, one usually cuts off the kinetic description at a momentum scale of the order of the Debye screening mass (see e.g. \cite{York:2014wja}). 

However an equation such as Eq.~(\ref{transport0}) may have a wider range of validity than that suggested by the weak coupling analysis. There are situations where non linear field equations can be well approximated by kinetic theory, interpreting the distribution function as the amplitude of a classical wave. The non linear interactions are reduced to the most relevant ones, in the present case, quartic and trilinear coupling, with given strength, the effects of  higher non linear couplings being simply to renormalize  the lowest order ones.  
One could also argue that the kinetic equations that we are using have a finite limit when the Debye screening mass and other thermal masses are sent to zero, except in places where they play a crucial role in fixing the strength of the various interactions.  In this limit the equations simplify, to the point where analytic results can be obtained.  We may add  that we  have explored the effects of   finite masses, and observed no major qualitative changes, while these finite masses make the calculations much more complicated and hence less transparent (see in particular \cite{Blaizot:2015wga} for a detailed study in the case of elastic scattering).

The outline of the paper is as follows. After the next section devoted to general considerations on the thermalization, and simple analytical estimates, we proceed in sections 3 and 4 to the analysis of thermalization when, respectively, only elastic or inelastic processes are present. We shall discover strong similarities between these two cases, in  spite of the obvious differences in the physical processes involved. In the Sect. 5, we analyze the competition between elastic and inelastic processes, in the case where both have comparable rates. In particular, we investigate the role of inelastic processes in preventing the formation of a Bose-Einstein condensate. 
Then we conclude. In this paper, we focus on the small  momentum region.
A forthcoming publication will deal with other aspects of thermalization, in particular the large momentum region, as well as the longitudinal expansion. Finally, let us mention that a preliminary account of the present work was presented in Ref.~\cite{Blaizot:2016bsg}.

\section{General setting}\label{sec:general}
We  describe the system of gluons produced in an ultra-relativistic heavy ion collision by a distribution function in phase-space, $f(\x,\p)$, which represents  the number of gluons of a given spin and color in the phase-space element $d^3\x d^3\p/(2\pi)^3$. The present discussion is limited to the situation where $f(\x,\p)$ is uniform in space, i.e., independent of $\x$. It is also independent of spin and color, and finally we assume that it is isotropic in momentum space. Thus, it depends only on the magnitude of the 3-momentum, $p=|\p|$. It also depends on time, and its evolution is  obtained by solving the kinetic equation (\ref{transport0}).

\begin{figure}[!hbt]
\begin{center}
\includegraphics[width=0.5\textwidth]{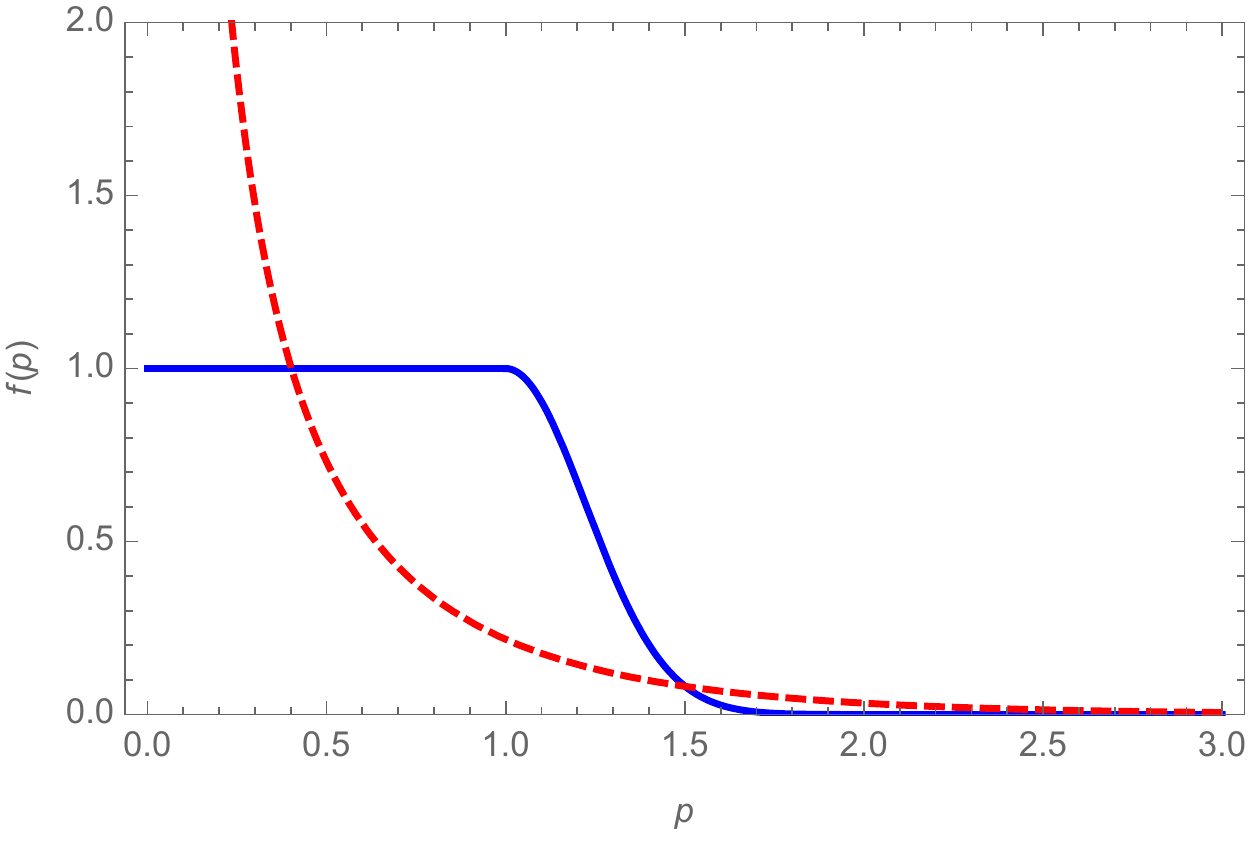}\includegraphics[width=0.5\textwidth]{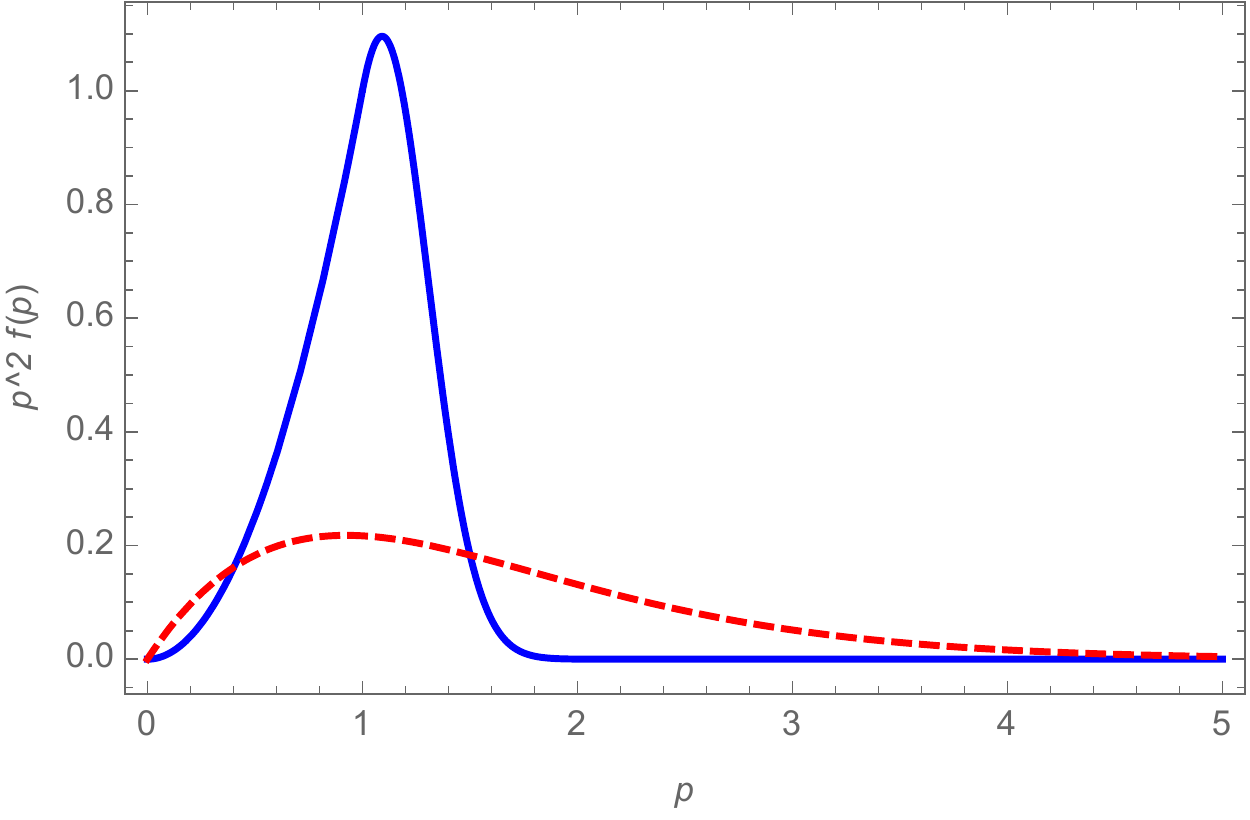}
\caption{(Color online). This figure shows the generic transformation of the distribution function $f(t,p)$ as the system evolves from the initial state, represented by the full (blue) line to the equilibrium distribution represented by the dashed (red) line. Left: the distribution itself. Right: the distribution multiplied by the phase space factor $p^2$. The initial distribution is the smooth distribution of Eq.~(\ref{eq_glasma_fs}). } \vspace{-0.25in}
\label{fig:fig0}
\end{center}
\end{figure}

We consider a family of initial conditions of the generic form $f(p)=f_0\, g(p/Q_s) $, where $g(x)$ is a dimensionless function. Such initial distributions are inspired by the color glass (CGC)  picture \cite{Blaizot:2011xf}. They are characterized by two parameters:  $Q_s$, the saturation momentum,  which  fixes the scale of momenta, and $f_0$, a typical occupation factor. A particularly simple distribution is 
\begin{eqnarray} \label{eq_glasma_f}
f( 0,p ) = f_0 \, \theta(1 - p/Q_s ). 
\end{eqnarray}
It assumes that all gluons with $p\lesssim Q_s$ are freed in the collision, and have initially the same occupation $f_0$. This distribution will be used below for simple analytical estimates. However, in the numerical calculations to be presented in this paper, we have used  a smooth version of this distribution, namely:
\beq\label{eq_glasma_fs}
f(0,p) = f_0 \,  \Theta(1-p/Q_s)  + f_0  \,  \Theta(p/Q_s -1 ) \,  \rme^{-10 \left( p/Q_s-1\right)^2}.
\eeq
Note that this smooth distribution has a ``tail'' extending up to momenta $p\simeq 1.5 \,Q_s$ (see Fig.~\ref{fig:fig0}).

The final state is given by a Bose distribution of the form
\beq\label{Bose-equil}
f_{\rm eq}(p)=\frac{1}{\rme^{(p-\mu_{\rm eq})/T_{\rm eq}}-1},
\eeq
with $T_{\rm eq}$ and $\mu_{\rm eq}$ respectively the equilibrium temperature and chemical potential (we ignore quarks in the present discussion). The presence of the chemical potential reflects the conservation of particle number and is of course non vanishing only in the case where number changing processes can be neglected, that is in the case where $C_{\rm inel}$ is set to zero in Eq.~(\ref{transport0}). In such a case, the equilibrium chemical potential will be either negative (underpopulation) or zero (overpopulation), with, in the latter case, formation of a condensate \cite{Blaizot:2011xf,Blaizot:2013lga} 
to accommodate the particles that do not fit in the distribution (\ref{Bose-equil}).  In the case where inelastic collisions cannot be ignored, the equilibrium distribution always corresponds to a vanishing chemical potential. Then the equilibrium distribution is completely determined by the temperature, which is fixed by the initial energy.

 We now focus on such systems for which the chemical potential vanishes in equilibrium.  For a CGC type initial condition, Eq.~(\ref{eq_glasma_f}),  the initial energy density is  $
 \epsilon_{\rm in} \equiv \int \rmd  p \,p^3 f(0,p)/(4\pi) ={f_0 Q_s^4}/{(8\pi^2)}$.
 By comparing with a thermal  distribution $\epsilon_{\rm in}=\epsilon_{\rm eq}=T_{\rm eq}^4{\pi^2}/{30}$, one obtains the equilibrium temperature
 \begin{eqnarray}\label{Tequilibrium}
 T_{\rm eq} = \left(\frac{15 f_0}{4\pi^4}\right)^{1/4} Q_s \approx 0.44 \, f_0^{1/4} Q_s.
 \end{eqnarray}
 For realistic values of $f_0$, this temperature is less than $Q_s$. It is only for very large overpopulation, i.e., 
 when $f_0\gtrsim 26$,  that $T_{\rm eq} \gtrsim Q_s$. 
 
 With the same initial distribution, the initial particle number density is 
  \begin{eqnarray}
 n_{\rm in} = \frac{f_0 Q_s^3}{6\pi^2},
 \end{eqnarray}
 while in thermal equilibrium at temperature $T_{\rm eq} $ it is 
$ n_{\rm eq} = {\zeta(3) T_{\rm eq}^3}/{\pi^2}  \propto f_0^{3/4}$,
 so that  
 \begin{eqnarray}
\frac{n_{\rm eq}}{n_{\rm in}} = \frac{6 \zeta(3) (15/4)^{3/4}}{f_0^{1/4} \pi^3} .
 \end{eqnarray}
 This  ratio is unity when $f_0 = f_c \approx 0.154$. If $f_0>f_c$ the system is overpopulated and the total particle number decreases during equilibration, while if $f_0<f_c $  the system is underpopulated and the total particle number  increases via inelastic processes in order to reach equilibrium. Similarly, the energy per particle in the initial distribution is $\epsilon_{\rm in}/n_{\rm in}=0.75\,Q_s $, while it is $\epsilon_{\rm eq}/n_{\rm eq}=T_{\rm eq}\pi^4/(30\zeta(3))\approx 1.18 \, f_0^{1/4}\, Q_s$,  the two quantities coinciding for $f_0=f_c$.  Note that the critical value of $f_0$ is independent of whether the system is driven to equilibrium by elastic or inelastic processes: it only depends on the fact that the chemical potential vanishes in equilibrium. Note also that the numerical value of $f_0$ depends somewhat on the initial density profile, that is on the function $g(p/Q_s)$. Thus, for the smooth distribution (\ref{eq_glasma_fs}), $f_c=0.167$.\\

The main question that we want to address in this paper is illustrated by Fig.~\ref{fig:fig0}, where the initial and final distributions are displayed for a value $f_0=1$ corresponding to an overpopulated initial sate. We want to understand how the system evolves in order to reach the final distribution, focussing more particularly on the soft momentum region. Note that this region represents a small fraction of the total phase space, and concerns a small number of particles. As clearly shown by the right panel of Fig.~\ref{fig:fig0}, the total particle density is indeed dominated by the hard particles, with momenta of order $Q_s$. Still, as we shall see, the evolution of the soft sector involves interesting physics that play an important role in the whole process of thermalization. It determines in particular the way the excess particles are eliminated.

A number of qualitative features can be inferred from Fig.~\ref{fig:fig0}, and elementary estimates, without actually solving the kinetic equation. 
 The initial distribution plotted in Fig.~\ref{fig:fig0} corresponds to $f_0=1$, but the main qualitative features are generic of the situation with overpopulation, and the chosen initial profile. One observes in particular: i) a decrease of the total density (more clearly visible in the right panel)-- with the initial profile (\ref{eq_glasma_fs}) we have $n_{\rm eq}=0.64\, n_{\rm in}$; ii) a small increase of particle number in the soft region; iii) the population of the high momentum tail of the distribution. 
 
 In order to characterize more quantitatively the soft and hard regions, we note that there are two points where the final distribution crosses the initial one. Consider first  $f_{\rm eq}(p=Q_s)$. This is given by  
 \begin{eqnarray}
 f_{\rm eq} (p=Q_s) = \frac{1}{e^{Q_s/T_{\rm eq}}-1} < f_0 .
 \end{eqnarray}
 It can be indeed verified that, with the value of $T_{\rm eq}$ given above, Eq.~(\ref{Tequilibrium}), this inequality is satisfied for any value of $f_0$. That is, the value of $f$ at $p=Q_s$ will always decrease during thermalization, irrespective of the initial value $f_0$: the particles in the vicinity of $Q_s$ are pushed to higher momenta in order to populate the tail of the equilibrium distribution. For instance, with the initial profile (\ref{eq_glasma_fs}), the density of particles with momenta $p\gtrsim Q_s$ goes from $0.54 \, n_{\rm in}$ initially to $0.78\, n_{\rm in}$ in equilibrium. Similarly, one can  define the energy $\bar p$ at which $f_{\rm eq}(\bar p)=f_0$. This is given by  
 \begin{eqnarray}\label{baromega}
 \bar p = T_{\rm eq} \ln\left(1+\frac{1}{f_0}\right) < Q_s,
 \end{eqnarray}
 where again the inequality follows from the relation (\ref{Tequilibrium}). This momentum $\bar p$ may be taken as a measure of the extent of the ``soft region".  Note that, as can be guessed from  the plot in the right panel of Fig.~\ref{fig:fig0},  the total number of particles in the soft region is very small, and increases by a tiny amount: for the initial profile (\ref{eq_glasma_fs}), $\bar p\simeq 0.31$,  and the density of particles with momenta $p\lesssim \bar p$ goes from $3\times 10^{-3}\, n_{\rm in}$ initially to  $5\times 10^{-3}\, n_{\rm in}$ in equilibrium.\\
 
Most of the numerical results to be presented in the next section correspond to the overpopulated situation with $f_0=1$. The corresponding initial energy and particle number densities are $\epsilon_{\rm in}\simeq 0.037$ and $n_{\rm in}\simeq 0.037$. The equilibrium temperature is $T_{\rm eq}\simeq 0.58$ and the equilibrium  number density is $n_{\rm eq}\simeq 0.024$. Let us also indicate here, for completeness,  the values of other quantities that will be defined shortly (see Eqs.~(\ref{definitionsI}) and (\ref{Tstar0}) below): $I_a\simeq 0.067$, $I_b\simeq 0.084$ and $T_\ast\simeq 0.79$.

\section{Elastic scattering alone}\label{sec:elastic}

The approach to equilibrium when only $2\to 2$ scattering are involved has already been considered in previous studies using the Boltzmann kinetic equation in the small scattering angle approximation \cite{Mueller:1999pi,Blaizot:2013lga,Blaizot:2014jna}. In particular, it has been shown that there exist  two types of solution, according to whether $f_0$ is smaller or larger than $f_c$. For the underpopulated case, $f_0<f_c$, the solution evolves smoothly towards a Bose-Einstein  distribution, while for the overpopulated case, $f_0>f_c$, the system undergoes Bose-Einstein condensation (BEC) in order to eliminate the excess particles from the spectrum.\footnote{ Very similar equations have been considered long ago in another context, that of photons in equilibrium with a dilute gas of electrons. See for instance  \cite{Kompaneets1957,Zeldovich1969}.} 

In the small scattering angle approximation, the kinetic equation reduces to a Fokker-Planck equation of the form (see e.g. \cite{Blaizot:2013lga})
\beq\label{FPeqn}
\frac{\del f(t,p)}{\del t}=-\frac{1}{p^2}\frac{\del}{\del p}\left( p^2 {\mathcal J}(t,p)  \right).
\eeq
The right hand side is the divergence of a current (in momentum space), whose radial component is given in terms of the distribution function by
\beq\label{calJp0}
{\mathcal J}=-  4\pi\alpha^2N_c^2{\mathcal L}\left[ I_a\, \partial_p f + I_b\, f(1+f)\right] , \qquad {\cal F}=4\pi p^2 {\cal J},
\eeq
and ${\cal F}$ is the corresponding flux through a sphere of radius $p$.
In this expression,   ${\mathcal L}=\int \rmd q/q$ is the Coulomb logarithm, treated here as a constant of order unity, and $I_a$ and $I_b$ are the following integrals of the distribution function
\beq\label{definitionsI}
I_a\equiv \int\frac{d^3 p}{(2\pi)^3} f(p)(1+f(p)),\qquad I_b\equiv \int\frac{d^3 p}{(2\pi)^3}\frac{2f(p)}{p}.
\eeq 
For the initial distribution (\ref{eq_glasma_f}), these integrals are given by
\beq
I_a^0=\frac{Q_s^3}{6\pi^2} f_0(1+f_0),\qquad I_b^0=\frac{Q_s^2}{2\pi^2}f_0.
\eeq
As we shall recall shortly, the ratio 
\beq\label{eq:eff-temperature}
T^* \equiv  I_a/I_b,
\eeq
plays the role of an effective temperature for the soft region, which thermalizes rapidly. It equals the equilibrium temperature when the system is fully thermalized. At the initial time, 
\beq\label{Tstar0}
T^*_0= \frac{I_a^0}{I_b^0}=\frac{1+f_0}{3} Q_s.
\eeq
Note that $T^*_0$ grows linearly with $f_0$ while $T_{\rm eq}$ grows only as $f_0^{1/4}$ (see Eq.~(\ref{Tequilibrium})). We have 
\beq
\frac{T^*_0}{T_{\rm eq}}= \frac{\pi}{3}\left( \frac{4}{15} \right)^{1/4} \frac{1+f_0}{f_0^{1/4}}\approx 0.75\, \frac{1+f_0}{f_0^{1/4}}.
\eeq
This function of $f_0$ has a minimum larger than 1 ($\sim 1.33$)  for $f_0=1/3$. Therefore the effective temperature $T^*$ decreases when the system thermalizes, irrespective of the value of $f_0$.

It is convenient to redefine the time and set 
\beq\label{timescaling}
 \tau\equiv 4\pi^3 \bar\alpha^2 {\cal L}\;t ,
\eeq
where $\bar\alpha\equiv \alpha N_c/\pi$.\footnote{Taking ${\cal L}$ of order unity, $N_c=3$ and $\alpha=0.3$, one gets $36\pi\alpha^2{\cal L}\approx 10$ for the conversion factor between the physical time $t $ and $\tau$, i.e., $t\approx \tau/10$ with both $t $ and $\tau $ expressed in units of $Q_s^{-1}$. }
  After this rescaling, and measuring  the time $\tau$ in units of $Q_s^{-1}$ and momentum in units of $Q_s$, the transport equation contains no parameters. It is a universal equation, whose solutions are entirely determined by the initial conditions. This equation reads
  \beq\label{FPeqntau}
 \frac{\del f(\tau,p)}{\del \tau}=\frac{1}{p^2}\frac{\del}{\del p}\,  p^2 \left[  I_a\, \frac{\del}{\del p} f + I_b\, f(1+f)   \right] ,
\eeq
with the current simply given by 
\beq\label{calJp0b}
{\mathcal J}=- \left[ I_a\, \partial_p f + I_b\, f(1+f)\right] , \qquad {\cal F}=4\pi p^2 {\cal J}. 
\eeq

  The non-local character of the kinetic equation is worth-emphasizing: although Eq.~(\ref{FPeqn}) looks like a local partial differential equation for the function $f(\tau,p)$, there is in fact a non-linear coupling with the entire solution through the integrals $ I_a$ and $ I_b$ which enter as coefficients of the equation. Except perhaps in early transient regimes, these integrals may be viewed as ``slow variables'': they are dominated by hard particles, and their values adjust over a time scale that is large compared to that with governs the local variation of the distribution $f(p)$.  The integral $I_a$ is naturally proportional to a diffusion constant (in momentum space). It enters the expression of the so-called jet-quenching parameter $\hat q$ \cite{Baier:1996sk},  $\hat q=16 \pi^3\bar\alpha^2{\cal L} I_a$.  
 The integral  $I_b$ enters the definition of  the Debye screening mass, $m_D^2=2g^2N_cI_b$, although it is not immediately clear why the Debye mass should be related to the second piece of the current (\ref{calJp0b}). 
 
 In fact, additional insight into the physical significance of the two components of the current of Eq.~(\ref{calJp0}) that are proportional respectively to $I_a$ and $I_b$ can be gained by returning briefly to the derivation of the small angle approximation to the collision integral. We only need to consider  the statistical factors in the process $1+2\rightarrow 3+4$, where $\p_3=\p_1+\q$, $\p_4=\p_2-\q$, with $\q$ the momentum transfer. For the loss term, we have (with the shorthand notation $f_1=f_{\p_1}$, etc.)
 \beq
 A&=&f_1f_2 (1+f_{\p_1+\q})(1+f_{\p_2-\q})
\nn
&\approx& f_{1}f_{2} (1+f_{1})(1+f_{2}) +f_{2} (1+f_{2}) f_{1}\, \q\cdot\nab f_{1}-f_{1} (1+f_{1}) f_{2}\, \q\cdot\nab f_{2},\nn
 \eeq
 and for the gain term
 \beq
 -B&=&f_{\p_1+\q}f_{\p_2-\q} (1+f_1) (1+f_2)\nn
 &\approx& f_1f_2 (1+f_1)(1+f_2)+(1+f_1)(1+f_2) \left[ f_2\, \q\cdot\nab f_1-f_1\, \q\cdot\nab f_2\right].\nn
 \eeq
 Adding $A$ and $B$, we get two contributions. These are no longer associated to gain or loss terms, but  together they yield the two contributions to the current which are proportional to $I_a$ and $I_b$, respectively:
 \beq\label{A+B}
 A+B=-f_2 (1+f_2) \,\q\cdot\nab f_1+f_1(1+f_1) \,\q\cdot\nab f_2.
 \eeq
 The current ${\cal J}(p_1)$  is obtained by integrating over particle 2 (as well as over $\q$). After this integration, the first contribution in Eq.~(\ref{A+B}) represents the global effect of the scatterings of particle 1 on all the particles of the system. This contribution is proportional to the gradient of $f_1$, hence its interpretation as diffusion. The other contribution, proportional to the gradient of $f_2$, has in fact the same origin, and may be associated also with diffusion, or random walk. However the diffusion of  particles 2 induces the recoil of particle 1. It is then felt by particle 1 as a drag, opposing its own diffusion. This is the origin of the sign difference between the two components of the current and also of the terminology that we shall use in referring to the term proportional to $I_a$ as a diffusion current and to that proportional to $I_b$ as a drag current:
 \beq\label{calJdiffdrag}
{\mathcal J}_{\rm diff}=-   I_a\, \partial_p f ,\qquad  {\mathcal J}_{\rm drag}=-I_b\, f(1+f).
\eeq
  Note that the drag current plays an important role in carrying particles towards small momenta. The equilibrium results from a balance between these two components of the current, which cancel each other for a Bose-Einstein equilibrium distribution function $f_{\rm eq}(p)$ with $T_{\rm eq}=I_a/I_b$.

Conservation laws play an important role in the dynamics described by the Fokker-Planck  equation, Eq.~(\ref{FPeqntau}). The conservation of the particle number follows immediately from the fact that the collision integral is a divergence. When condensation occurs, the decrease of the number of particles in  the spectrum (i.e. those with momentum $p>0$) equals the flux of particles into the condensate at $p=0$. Energy conservation implies the following equality
\beq
I_a \int_\p \frac{\del f}{\del p}=I_b \int_\p f(1+f),
\eeq
which is automatically satisfied if $I_a$ and $I_b$ are self-consistently calculated from Eqs.~(\ref{definitionsI})\footnote{Note that we could consider a related Fokker-Planck equation in which $I_a$ and $I_b$ are held fixed. Then energy would not be conserved. However a thermal fixed point would still exist, with temperature $T=I_a/I_b$.}.\\
 
 We  now recall a few characteristic features of  thermalization when only elastic collisions are involved \cite{Blaizot:2013lga}. In the initial state, the drag current  dominates at small momentum, as is obvious from the expression (\ref{calJp0}) of the current, and the specific form of the initial distribution, Eq.~(\ref{eq_glasma_f}). However, this initial situation is unstable. Indeed a simple arguments shows that if the distribution is regular at the origin, i.e., if $f(p=0)$ is finite, the total current should vanish linearly with $p$. This result is obtained by integrating the kinetic equation in a small sphere of radius $p_0$. One then obtains $\dot f(0) p_0^3\sim p_0^2 {\cal J}(p_0)$ (with $\dot f=\rmd f/\rmd\tau$) , that is, ${\cal J}(p_0)\sim p_0$ as announced. 
  There is therefore initially a quasi instantaneous readjustment of the distribution function in order to guarantee this property \cite{Blaizot:2013lga}. 
  
  How this readjustment occurs, that is, how the linear behavior of the current  develops from an initial constant (momentum independent) current, is actually subtle. With the initial current constant, the kinetic equation yields, after an infinitesimal time step $\rmd t$, $\rmd f\sim \rmd t/p$. This suggests that a singular behavior $f(p)\sim 1/p$ immediately develops in the distribution function.  This is for instance what one observes by solving the Burgers equation, which is obtained by ignoring the diffusion current (see Appendix~\ref{sec:burgers}). In this case,  one indeed observes that condensation sets in immediately. However, the diffusion current changes the picture completely. Indeed it  cancels the contribution of the drag, bringing the system locally in  equilibrium, in essentially no time, with $f(p)$ taking approximately the form of a classical equilibrium distribution, $f(p)\simeq T^\ast/(p-\mu^\ast)$ with $T^\ast=I_a/I_b$, and $\mu^*$ related to $f(0)$.   
 
 \begin{figure}[!hbt]
\begin{center}
\includegraphics[width=0.7\textwidth]{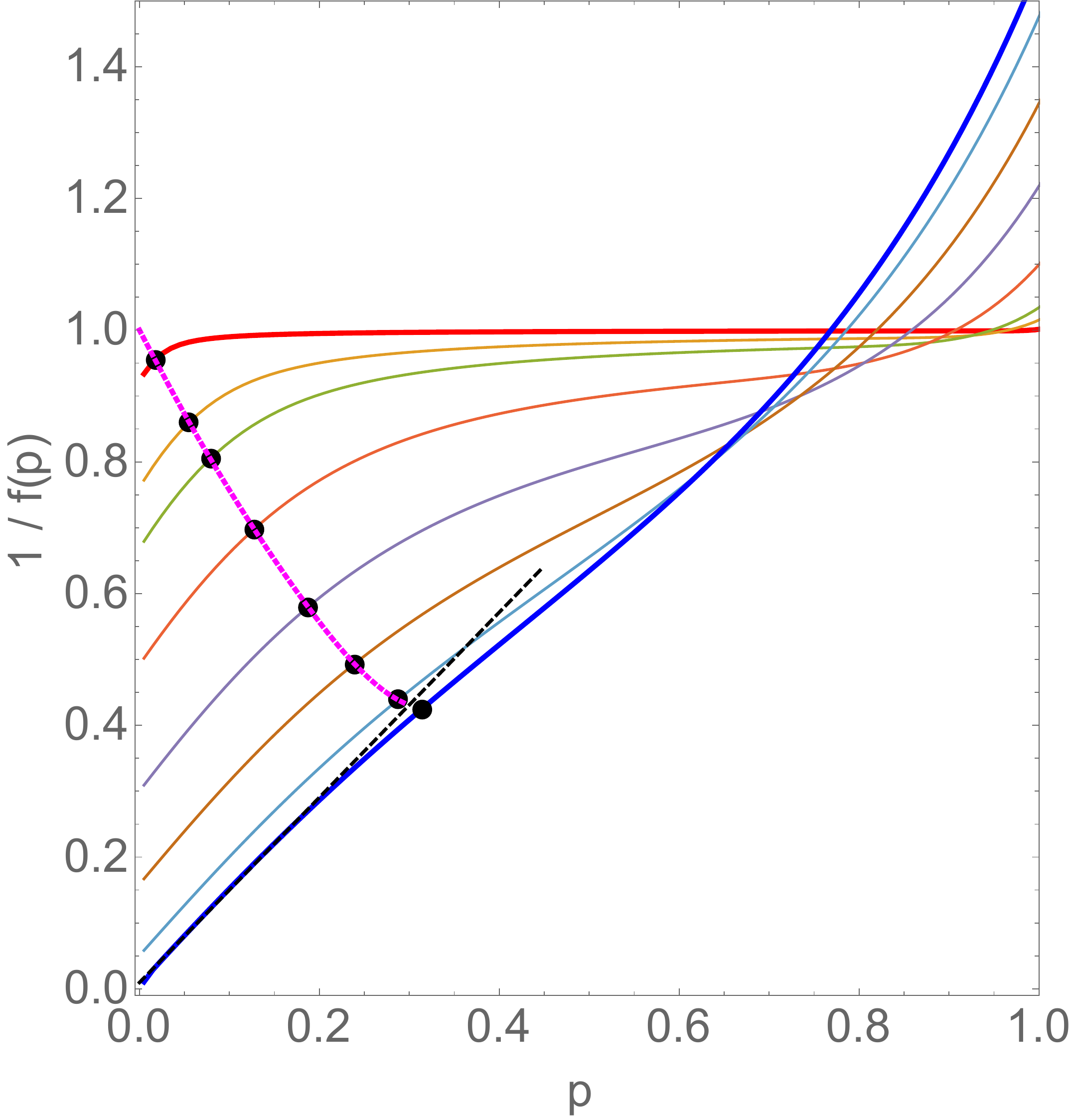}
\caption{(Color online). The inverse of the distribution function as a function of time. The times corresponding to the various plots are $\tau=$ 0.003, 0.030, 0.062, 0.162, 0.350, 0.568, 0.822, 0.982. The last curve (blue) corresponds to the onset for BEC, and the slope of the dashed line is determined by the local temperature $T^*$. Note that at onset a significant fraction of the soft region $p\lesssim 0.3 \,Q_s$ is thermalized. The dotted (magenta) line joining the black dots reveals the expansion of the soft region: the abscissa of  the black dots are obtained from the equation $p_\ast(\tau)=0.5 \sqrt{6 I_a \tau}$, where the factor 0.5 is a normalization factor adjusted so that $p_\ast(\tau=0.982)\simeq 0.3$. } \vspace{-0.25in}
\label{fig:finverse}
\end{center}
\end{figure}

 This argument reveals also a subtle aspect of BEC: although the initial conditions seem to favor the possibility of the instantaneous establishment of a constant flux at $p=0$, the system does not condense immediately. In fact the early time evolution of the system that we have just described  is independent of whether the system is under or overpopulated. This is natural, as it takes time for the system to ``know'' whether it is under or overpopulated. At early time, we are dealing with a local property of the kinetic equation, with a fast adjustment of the distribution. This adjustment leads to a fast equilibration of the softer part of the distribution, through the mutual cancellation of the drag and the diffusion currents, and the appearance of a chemical potential. The emergence of an effective chemical potential is natural in view of particle number conservation. However, this alone is not sufficient. For instance, no chemical potential appears in the solution of the Burgers equation, which also conserves particle number. For the chemical potential to develop, one needs in addition the presence of the thermal fixed point,  which requires both drag and diffusion. 
 
\begin{figure}[!hbt]
\begin{center}
\includegraphics[width=0.5\textwidth]{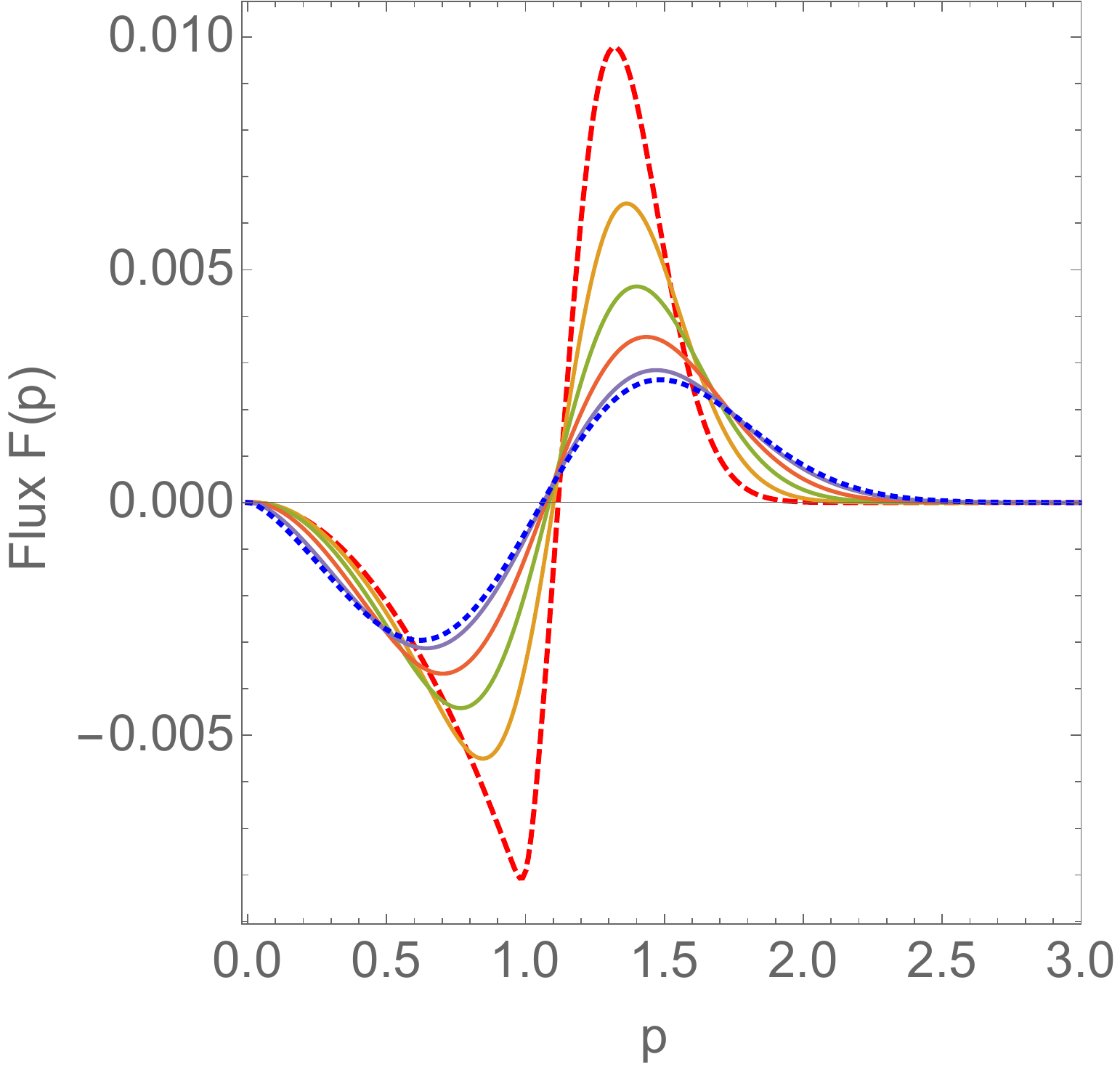}\includegraphics[width=0.5\textwidth]{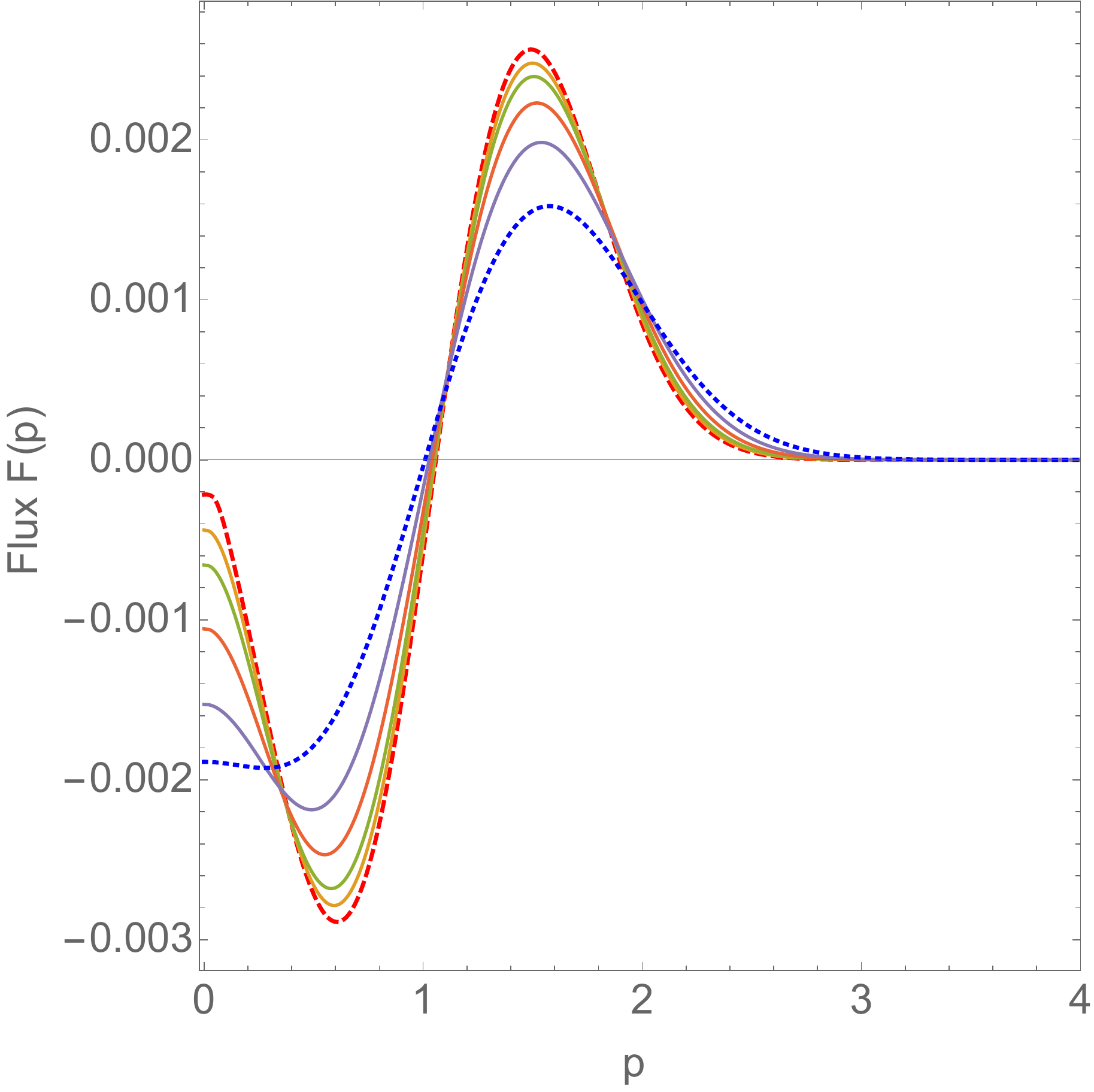}
\caption{(Color online). This figure shows the behavior of the flux as a function of momentum, before (left panel) and after (right panel) the onset of Bose-Einstein condensation.  On the left one sees for $p\lesssim 0.3 \,Q_s$  the development of the linear flux (blue curve) as onset is approached. This linear behavior is related \cite{Blaizot:2013lga} to the $1/p$ behavior of the current for $p\gtrsim \mu^*$. After onset, the distribution function near $p=0$ behaves as $1/p$, the current as $1/p^2$, and there is a constant flux of particles at the origin. The curves correspond to the following values of $\tau$: $\tau = 0.003, 0.162, 0.350, 0.568, 0.822, 0.935$ (left panel, from largest (red) to smallest (blue) amplitude), and $\tau = 0.994, 1.054, 1.117, 1.248, 1.460, 1.858$ (right panel). } \vspace{-0.25in}
\label{fig:fluxp}
\end{center}
\end{figure}

Now, for the classical thermal distribution, the total current vanishes, as is readily verified from Eq.~(\ref{calJp0}). There is however, at small $p$ ($p\lesssim |\mu^*|$), a small correction of the form $\delta f(p)=p^2 \dot f(0) /(6 I_a)$ which guarantees the linear dependence of the current mentioned above \cite{Blaizot:2013lga}. Note that this correction can be attributed to the diffusion term, after the largest part of the diffusion has cancelled the drag. The gradient of $\delta f$ is positive, corresponding to a negative diffusion current, i.e. to particles moving towards $p=0$,  gradually building the $1/p$ singularity that emerges as $\mu^*\to 0$. Similarly, at larger $p$ ($p\gtrsim |\mu^*|$),  the distribution function deviates from the classical distribution by a constant amount $\delta f(p)=-(1/2I_a) \rmd (T^*|\mu^*|)/\rmd\tau$. This can be associated in this case to a negative drag current ${\cal J}_{\rm drag}\sim (1/p) \rmd (T^*|\mu^*|)/\rmd\tau$.

 Because of the existence of these ``residual'' currents, after the initial stage particles continue to flow towards  small momenta. The region accurately described by the thermal distribution $T^\ast/(p-\mu^\ast)$ expands towards  larger momenta (see Fig.~\ref{fig:finverse}), the values of the local chemical potential $\mu^\ast$ and the temperature $T^\ast$ adjusting themselves as time goes on (generically $T^*$ decreases and $\mu^*$ increases). If the system is underpopulated, the evolution will continue smoothly until the entire distribution become a Bose-Einstein distribution, with chemical potential and temperature determined by the initial conditions. If the system is overpopulated, however, a point will be reached, in a finite time, where  the chemical potential vanishes: this marks the onset of BEC. At this point, a substantial fraction of the soft particles have thermalized.  As can be seen on Fig.~\ref{fig:finverse}, the distribution is well fitted by the classical thermal distribution up to momenta $p\lesssim 0.3 \, Q_s$.  As can also be seen on Fig.~\ref{fig:fluxp}, left panel, near onset, the current is dominated by the drag residual current, corresponding to a linear  flux.
 
 The growth of the soft momentum region, i.e., the region well fitted by the classical thermal distribution, can be understood from the correction to the distribution function at small $p$, namely $\delta f(p)\approx p^2 \dot f(0)/(6I_a)$. For a crude estimate, we  write $\dot f(0)\sim f(0)/\tau$, with $f(0)=T^*/|\mu^*|$, so that   $\delta f(p)/f(0)\approx p^2  /p_*^2(\tau)$ which exhibits a momentum scale  $p_\ast\propto \sqrt{6I_a\tau}$ at which the distribution function starts to deviate significantly from the classical distribution. One can recognize this scale $p_\ast$,  as well as its square root  dependence on ${\tau}$, in Fig.~\ref{fig:finverse}. The estimate just presented relies on an approximation valid when $p\lesssim |\mu^*|$. As $|\mu^*|$ decreases, the range of momenta where it applies shrinks, but one can verify on Fig.~\ref{fig:finverse} that $p_\ast$ remains smaller than $|\mu^*|$ down to small values of $|\mu^*|$ ($|\mu^*|\gtrsim 0.2$).

  The onset for BEC, corresponding to $|\mu^*|=0$,  is reached in a finite time $\tau_c$. At this time the soft region is thermalized, i.e, $p_\ast(\tau_c)\simeq \bar p$. The time $\tau_c$ was determined numerically in Ref.~\cite{Blaizot:2013lga} where it was shown that for $f_0\gtrsim 1$, $ \tau_c\simeq 2/{f_0(1+f_0)}$. This  estimate holds only for $f_0$ not too small ($f_0\gtrsim 1$), since when $f_0$ approaches the critical value $f_c$, the condensation time $\tau_c$ obviously becomes infinite. This finite time reflects the fact that it indeed takes time for the system to ``know" whether it is under or overpopulated, which requires the  soft sector to be thermalized. In the case of overpopulation, the drag current continues to push particles in the soft sector when this is already saturated, causing condensation. 
  
\begin{figure}[!hbt]
\begin{center}
\includegraphics[width=0.5\textwidth]{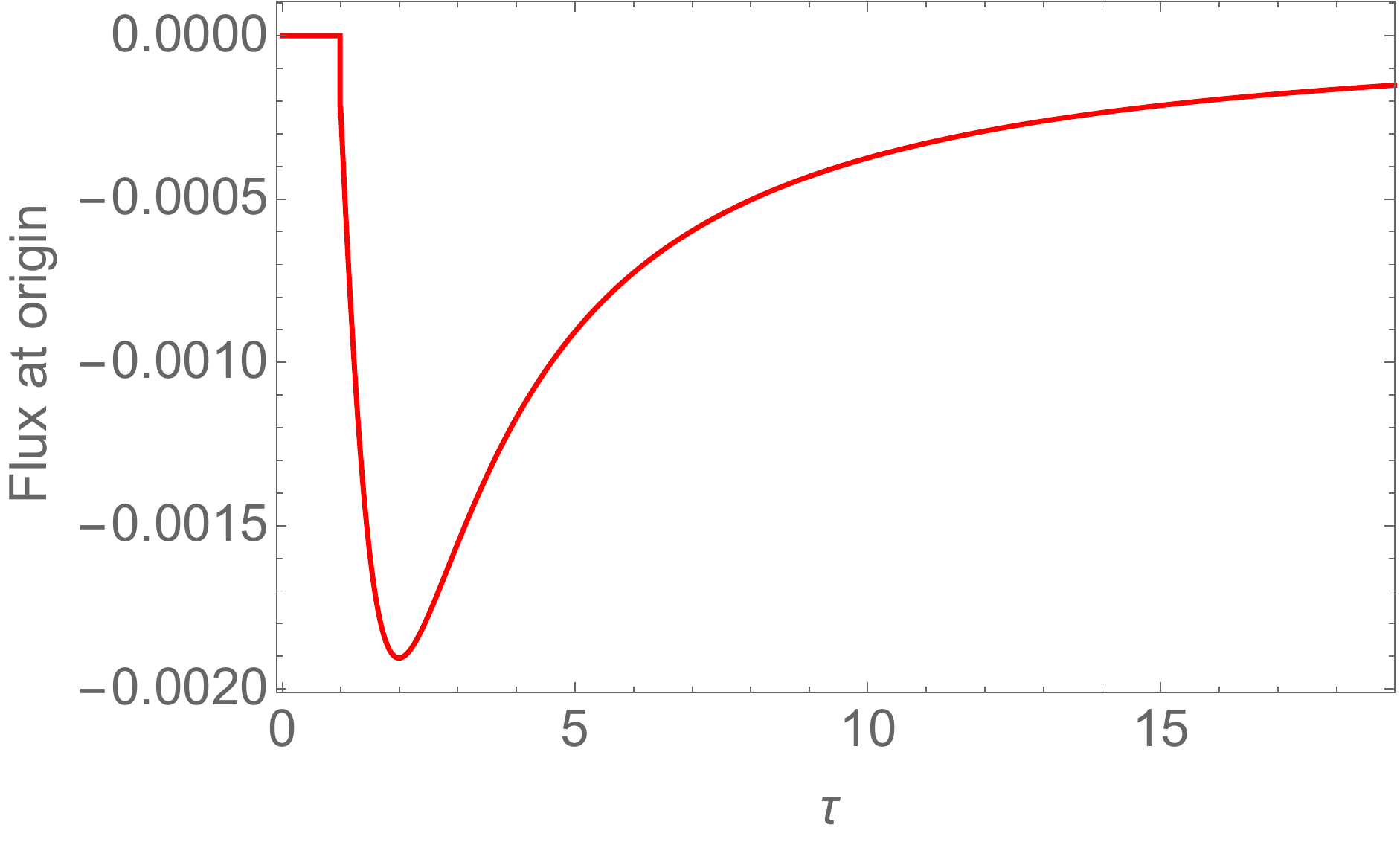}\includegraphics[width=0.5\textwidth]{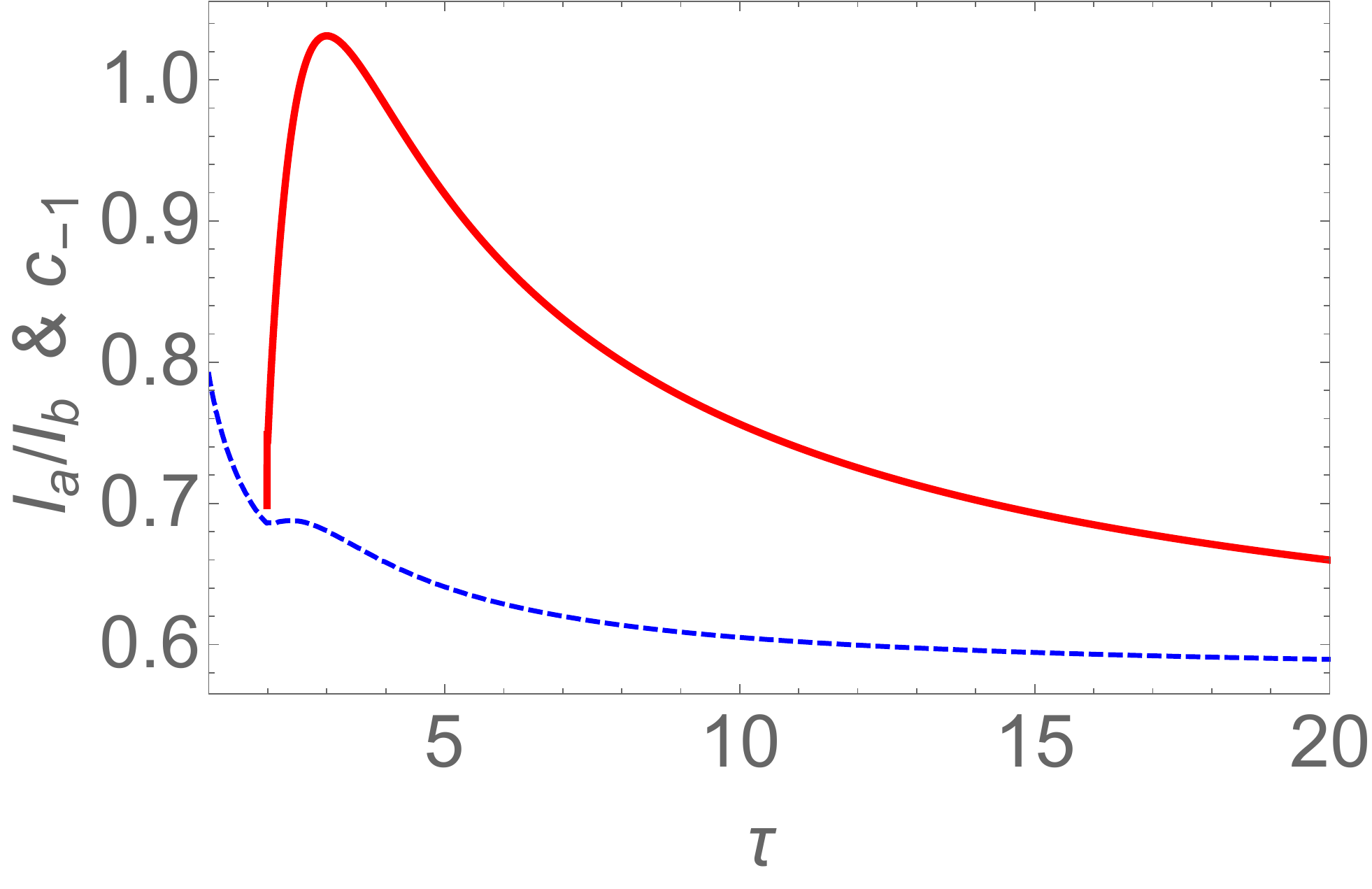}
\caption{(Color online). The left  figure shows the behavior of the flux at the origin, before onset of BEC where it vanishes, and after onset where it develops very rapidly. The threshold for condensation occurs at $\tau\approx 1$, at which time the density of non condensed particles starts to decrease. The right figure  illustrates the interplay between local  properties (the coefficient $c_{-1}$ of the distribution function) and global ones (the ratio of integrals $I_a/I_b=T^*$ represented by the blue dotted line) as a function of time. Before onset, $c_{-1}=0$. At the onset,  $\mu^*=0$ and $c_{-1}=T^*$. After the onset, particles start accumulating in the condensate, and the system is rapidly driven away from local equilibrium, with $c_{-1}$ (represented by the red curve) becoming significantly larger than the local effective temperature $T^*$. At later times, $c_{-1}$ decreases and eventually converges back to $T^*$. } \vspace{-0.25in}
\label{fig:fluxatorigin}
\end{center}
\end{figure}

  The equations that allow us to follow the system beyond the onset have recently been obtained in the small angle approximation \cite{Blaizot:2015xga}. We shall not use these equations here, but resort to a simpler procedure that consists  in switching from one solution to another at the onset, thereby ignoring the coupling between the thermal particles and the condensate (except for the trivial constraint of particle number conservation). The kinetic equation admits indeed two solutions which are distinguished by their behaviors at the origin  (see \cite{Blaizot:2014jna}, Appendix B). The first solution is regular at $p=0$, i.e., $f(p\to 0)$ is finite. This regular solution describes the thermalization of underpopulated systems. The other solution behaves as $c_{-1}/p-1/2+c_1 p+\cdots$ near $p=0$. This solution does not conserve particle number, but allows particles to ``disappear'' at $p=0$. That is, the complete solution can be written as $f(p)+(2\pi)^3n_c\delta^{(3)}(\p)$, with $n_c$ interpreted as the density of the condensate. The flux at the origin ($ {\cal F}(0)= {\cal F}(p=0)$) is related to the coefficient $c_{-1}$:
  \beq\label{flow0}
  {\cal F}(0)=4\pi I_a c_{-1}\left( 1-\frac{c_{-1}}{T^*}\right), \qquad 
c_{-1}=\frac{T_\ast}{2}\left( 1+\sqrt{1+\frac{{\cal F}(0)}{\pi I_a T_\ast}}\right).
  \eeq
  Note that at the onset of condensation, when we switch from one solution to the other, $c_{-1}=T^*$. Just after onset, $c_{-1}$ becomes larger than $T^*$, producing a negative flux of particles. The number of particles that flow in the condensate can be obtained simply by integrating this flux: $\dot n_c={\cal F}(0)/(2\pi)^3$. The behavior of the flux before and after onset is illustrated in Fig.~\ref{fig:fluxp}.

Finally Fig.~\ref{fig:fluxatorigin}, 
 illustrates the rapid development of the condensate beyond onset, and the global versus local aspects of the phenomenon. As we have already indicated, when the system is overpopulated, a strong drag current pushes particles to low momenta.  As soon as the onset is reached, the system is rapidly driven out of equilibrium and its behavior is largely determined by local properties, such as the local distribution function and the coefficient $c_{-1}$ of the $1/p$ term. This coefficient increases rapidly, as the flow of particles into the condensate develops. There is then a mismatch between the local effective temperature $T^*$ and the coefficient $c_{-1}$ that eventually disappears as the system approaches equilibrium and the flow vanishes (see Eq.~(\ref{flow0}). \\

  In summary, the elastic collisions drive the system to thermal equilibrium, starting with the very small momentum region. A ``soft region" develops, which is well described by a classical thermal distribution, with time dependent parameters $T^*$ and $\mu^*$. As times goes on, the soft region expands to larger momenta, the extent of this region growing as $p_\ast(\tau)\sim \sqrt{\tau}$. The chemical potential vanishes at some finite time $\tau_c$, at which point the entire soft region ($p\lesssim \bar p$) is thermalized. After $\tau_c$ the particle number in the spectrum starts to decrease, particles being pushed into the condensate by the drag current.

\section{Inelastic scattering alone} \label{sec:inel-scatt}

 We now  focus on the inelastic scattering, leaving temporarily aside the elastic collisions. Although it may look somewhat artificial to drop the elastic collisions since, as we shall recall shortly, the kind of inelastic processes that we shall consider are triggered by elastic collisions,  it is instructive to analyze how the inelastic scatterings alone drive the system to equilibrium. 
 
 The leading order inelastic processes are $2 \to 3$ processes (and their reverse) which involve the production of an extra gluon. The dominant contributions come from kinematical configurations where  the additional  gluon is emitted after a small angle scattering in a direction almost parallel to that of one of the scattered gluons.  Such processes are amplified by infrared divergences, which are naturally regulated by the thermal mass. As a result they become of the same order of magnitude as the corresponding elastic processes \cite{Baier:2000sb,Arnold:2002zm}. Furthermore, a factorization emerges in this regime, which allows to reduce the matrix element to an effective $1 \to 2$ matrix element. The resulting dynamics is then very similar to that involved in the gluon cascades that occur in jets (see e.g.  \cite{Blaizot:2015lma}, and references therein), and this analogy is used in Appendix~\ref{derivationkinetic} to provide a simple derivation of the corresponding kinetic equations. Detailed calculations of the  $2 \to 3$ processes in the relevant kinematical  regime are presented in Refs.~\cite{Huang:2013lia}  and \cite{yacine}.
 
\begin{figure}[!hbt]
\begin{center}
\includegraphics[width=0.8\textwidth]{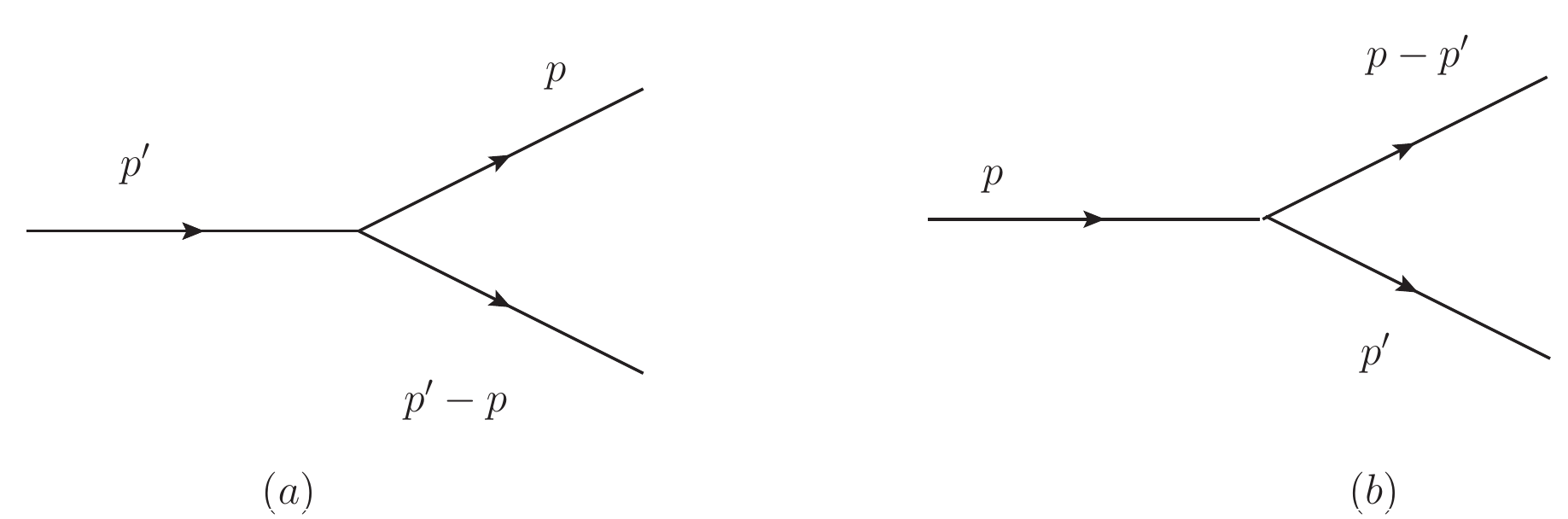}
\caption{Inelastic splitting processes. Left: gain term (process that populate the state with energy $p$). Right, loss term that depletes the state $p$. There are correspondingly two reverse processes, corresponding to merging, which are not drawn.  } \vspace{-0.25in}
\label{fig:splitting}
\end{center}
\end{figure}

The relevant kinetic equations are derived in Appendix~\ref{derivationkinetic}. We shall consider in this paper only the Bethe-Heitler (BH) regime, that is, we ignore possible interference effects that may take place among the multiple scattering leading to radiation (Landau-Pomeranchuk-Migdal (LPM)  effect). Since we are focussing on soft modes, the BH regime is the most relevant. The corresponding kinetic equation takes the form 
\beq
 \label{eq_D_BH_5}
\partial_\tau f(p)= \frac{RT^*}{ p^3}  \left\{ \int_0^\infty \rmd k K(p,p+k) \Phi(p,p+k) -\int_0^p\rmd k K(k,p) \Phi(k,p)\right\}.
\eeq
 where the (simplified) kernel $K(p,p')$ is 
 \beq\label{kernelKpp}
 K(p,p')=\frac{p'^3}{p'-p},
 \eeq
 and $\Phi(p,p')$ is the following combination of statistical factors
 \beq
\Phi(p,p')\equiv f(p')+f(p')f(p)+f(p'-p)\left[ f(p')-f(p)  \right].
\eeq
The kernel (\ref{kernelKpp}) results from an approximation valid when the inelastic processes are  strongly asymmetric, i.e., $|p-p^\prime|\ll p$, which are indeed the dominant processes. A further simplification made to get (\ref{kernelKpp}) involves dropping a smoothly varying function of $p$ and $p'$ to keep in the kernel only the dominant singular structure as $p-p'\to 0$. 
In arriving at Eq.~(\ref{eq_D_BH_5}), we have rescaled the time as in Eq.~(\ref{timescaling}). The quantity $R$ is a parameter, expected to be of order unity,  that controls the relative strength of elastic and inelastic scatterings \cite{Huang:2013lia}.  Its value in a strict weak coupling expansion is $R=1.83$ \cite{yacine}. However, since we are interested in identifying generic behaviors, and how these  may depend on the value of $R$, we keep $R$  as a free parameter (although most of the numerical results presented in this section and the next corresponds to $R=1$). Besides, several approximations have been made that may affect the value of  $R$.  For instance, in rescaling the time, we have ignored the Coulomb logarithm present in the elastic kernel. Also, when $2\to 3$ processes become important, so do other inelastic processes. To the extent that these higher order processes can be mimicked by effective $1\to 2$ processes, one could expect these to renormalize $R$. Such a renormalization is in fact already present in the  reduction of $2 \to 3$ processes, which contains a contribution that modifies the elastic kernel \cite{Huang:2013lia}.

The physical content of the kinetic equation (\ref{eq_D_BH_5}) can be analyzed in terms of splitting and merging processes, and gain and loss terms (see Fig.~\ref{fig:splitting}). Both splitting and merging contribute to the gain and the loss terms. Note that, generically, merging processes increase the energy of the particle that one follows, while splitting processes decrease it, as illustrated in Fig.~\ref{fig:gainloss}. Said differently, splitting processes move particles towards lower energy, while merging processes do the opposite. Thus,  splittings and mergings play similar roles as, respectively, the drag and diffusion currents in the elastic case. The equilibrium results from a balance between these two antagonistic processes.

\subsection{Approximate equation at small momenta}
 We first focus on the small $p$ region, and obtain a simplified version of the kinetic equation which accurately describes the infrared behavior of the system. As we shall see, the dominant behavior of $f(p)$ at small $p$ will turn out to be $f(p)\sim 1/p$. We then exploit this result to  estimate the various integrals, and verify a posteriori that the approximation to be derived is self-consistent. Let us consider the second integral in Eq.~(\ref{eq_D_BH_5}). In this integral, in the limit $p\to 0$, we may replace $K(k,p)\rightarrow k^2$, and $\Phi(k,p)\to 1/p$ (assuming that $f(k)\sim 1/k$ and $f(p)\sim 1/p$). It follows then that the integral vanishes as $p^2$ when $p\to 0$.  Consider next  the first integral. There too, we can replace $K(p,p+k)\to k^2$, and the integral is dominated by values $k\gg p$. This allows us to write\footnote{Similar approximations of the statistical factors are used in \cite{Mueller:2006up}.}
\beq
\Phi(p,p+k)\approx f(k)+p f(p) \frac{\del f}{\del k}+f(k)^2.
\eeq
Eq.~(\ref{eq_D_BH_5}) can then be written as
\beq\label{approxequation}
\partial_\tau f(p) &\approx&\frac{RT^*}{p^3} \left\{ p f(p)\int_{p_c}^\infty \rmd k k^2 \frac{\del f}{\del k} +\int_{p_c}^\infty \rmd k k^2 f(k)(1+f(k)) \right\}\nn
&\approx& \frac{RT^*}{p^3} \left[ I_a-I_b p f(p) \right]=\frac{RI_a}{p^3}\left[ T^*-pf(p)   \right],
\eeq
where we encounter again the integrals $I_a$ and $I_b$  introduced in the previous section on elastic scattering (cf. Eqs.~(\ref{definitionsI}) and (\ref{eq:eff-temperature})). In the first line, we have introduced a cutoff $p\ll p_c\lesssim k$ to control the approximation that we have done ($k\gg p$), but in the last line we have let $p_c\to 0$ since the integrals are dominated by large $k$ and are regular as $p_c\to 0$. Equation~(\ref{approxequation}) is a simple differential equation that controls accurately the small momentum behavior of the distribution. It is an approximate equation. In particular, in contrast to Eq.~(\ref{eq_D_BH_5}),  it does not conserve energy, and its fixed point is the classical thermal distribution $T^*/p$ rather than the Bose distribution. Thus, it should be used only for small times, and small momenta.

The equation (\ref{approxequation}) makes it clear that  in order to avoid the blow up of the solution, we must have,
 at small $p$,  
\beq
 f(p)=\frac{T^*}{p}, 
 \eeq
 as anticipated. 
In fact, the solution of the equation can be written as follows
 \beq\label{fullsolution}
 f(p,\tau)&=&f_0(p) \rme^{-\frac{R}{p^2}\int_0^\tau \rmd u I_a(u)} +\frac{R}{p^3} \int_0^\tau \rmd \tau' I_a(\tau') T^*(\tau') \rme^{-\frac{R}{p^2} \int_{\tau'}^\tau \rmd u I_a(u)}\nn
 &=& \left(f_0(p)-\frac{T^*(0)}{p}  \right)\rme^{-\frac{p_\ast^2(\tau)}{p^2}}+\frac{1}{p}\left(T^*(\tau)-\int_0^\tau \rmd\tau' \rme^{-\frac{p^2_\ast(\tau)-p^2_\ast(\tau')}{p^2}} \frac{\rmd T^*}{\rmd\tau'}   \right)\nn
 \eeq
 with 
 \beq
 p_\ast^2(\tau)\equiv R \int_0^\tau \rmd u I_a(u).
 \eeq
 Actually, this is not truly a ``solution'', since $I_a$ and $T^\ast$ in the right hand side of Eq.~(\ref{fullsolution}) depends on $f$ itself. However, this writing allows us to see that the qualitative behavior of the solution is dominated by an essential singularity at vanishing momentum. This singularity ensures that, at small $p$, $f(p,\tau)$ behaves as $T^*(\tau)/p$, as 
shown by the second line. The equation (\ref{fullsolution}) 
 can  be further  simplified  if one ignores the (weak) time dependence of $I_a$ and $T^*$, thereby getting a true solution:
 \beq\label{solutioninel}
f(p,\tau)&=&\frac{T^*}{p}+\left( f_0(p)- \frac{T^*}{p} \right)\rme^{-p_*^2(\tau)/p^2} ,
\eeq
with now $ p_*^2(\tau)=RI_a \tau$. As we shall verify shortly, this approximate solution captures the main qualitative (and semi-quantitative) features of Eq.~(\ref{approxequation}), this equation (\ref{approxequation}) being itself an accurate approximation to the full kinetic equation, Eq.~(\ref{eq_D_BH_5}), at small momenta. \\

The essential singularity at small momentum in Eq.~(\ref{solutioninel}) is controlled by the scale $p_\ast(\tau)$, which vanishes as $\sqrt{\tau}$ when  $\tau\to 0$ .\footnote{The existence of the two scales $T^*$ and $p^*$ characterizing the soft sector of the distribution function was also recognized in Refs.~\cite{Mueller:2006up,Kurkela:2011ti}.} Thus, for any finite $\tau$, $p_\ast(\tau)$ is non zero, and $\exp(-p_\ast^2/p^2)$ essentially vanishes for all $p\lesssim p_\ast$. In the region $p\lesssim p_\ast$,  $f(p,\tau)$ takes the form of a classical equilibrium distribution function, $f(p,\tau)\simeq T^\ast/p$. Physically, what happens is that the soft modes are rapidly populated by radiation from hard particles.  The spectrum of such radiated soft modes diverges at small $p$ as $1/p^3$, and immediately exceeds the equilibrium spectrum by a large amount. In this high density of soft modes, merging processes occur at a high rate and efficiently eliminate the soft modes in excess, thereby driving  the system to equilibrium. The essential singularity is the mathematical translation of the fact that, near $p=0$, this occurs in essentially no time.

\begin{figure}[!hbt]
\begin{center}
\includegraphics[width=0.75\textwidth]{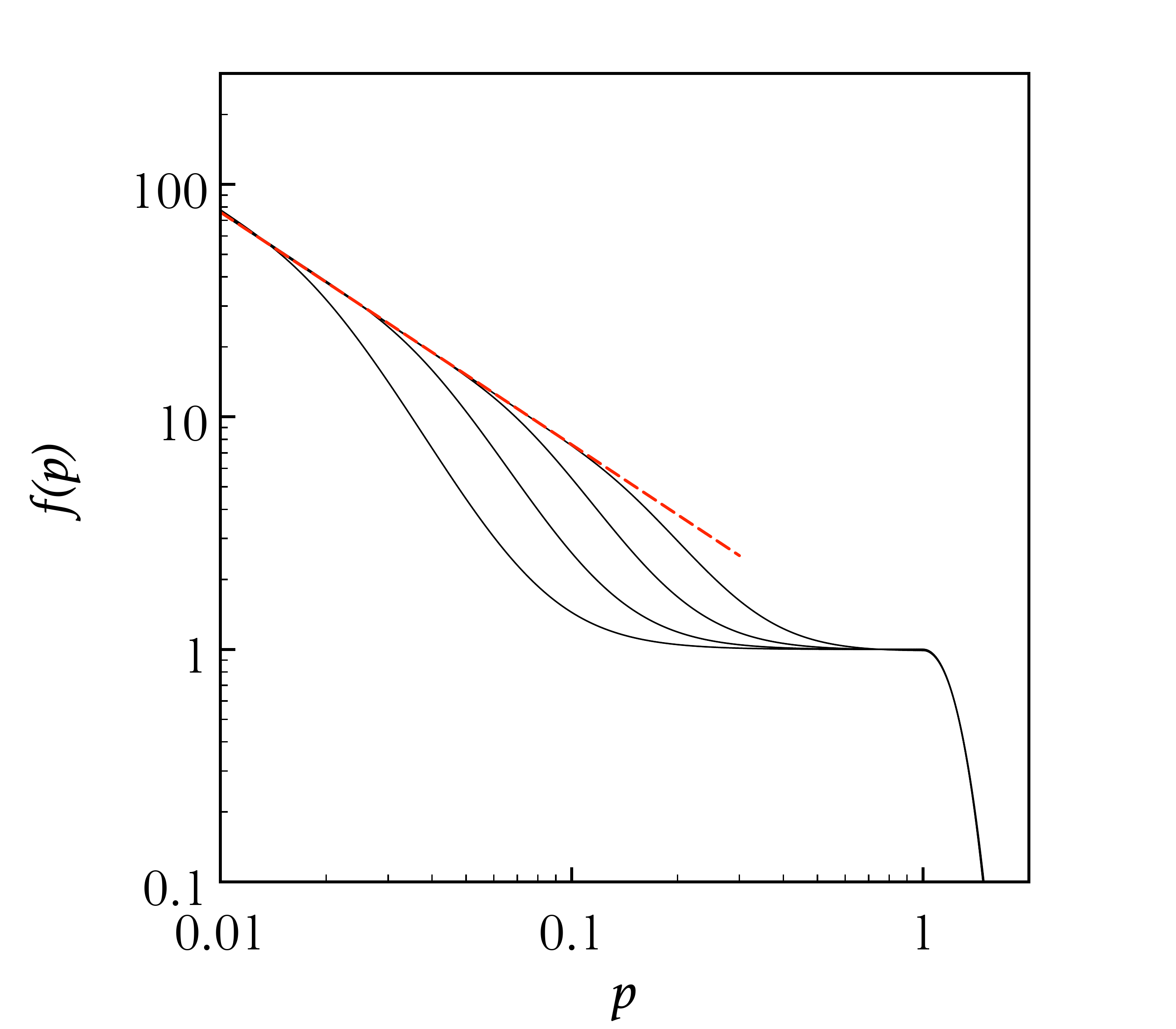}
 \vspace{-0.1in}
\caption{(Color online). The solution of the approximate equation  (\ref{approxequation}) at early times. From bottom to top: $\tau=$ 0.01, 0.04, 0.16, 0.64. The dashed red line corresponds to the  thermal distribution $T_\ast/p$, with $T_\ast=I_a/I_b=0.759$ at $\tau=0.64$.  }
\vspace{-0.25in}
\label{fig:approximate1}
\end{center}
\end{figure}

This can also be seen in another way. At a given momentum $p$, one may regard $\exp(-p_\ast^2/p^2)$ as a relaxation factor, $\exp(-p_\ast^2/p^2)=\exp(-\tau/\tau_\ast(p))$ with a momentum dependent relaxation time $\tau_\ast(p)=p^2/(RI_a)$. The relaxation process brings the initial distribution $f_0$ to the final equilibrium one, $T^*/p$, in a time scale $\tau_\ast\sim p^2$ which vanishes when $p=0$, reflecting the quasi instantaneous equilibration of the very soft modes, as already discussed.

The (numerical) solution of Eq.~(\ref{approxequation})  at early times is plotted in Fig.~\ref{fig:approximate1}. One sees clearly the development of the region where soft particles thermalize and where  $f(p,t)=T^\ast/p$. The size of this  region grows as $p_\ast(t)\sim \sqrt{\tau}$, while the temperature $T^\ast(\tau)$ remains approximately constant (this is so for $\tau\lesssim 1$). As equilibrium is being established in a small momentum region, hard particles continue to radiate soft particles, but these are quickly eliminated by the merging processes. This results in a  front located around  $p\sim p_\ast(\tau)$ which moves towards hard momenta. In the region of the front, the shape of the distribution  is determined by the radiation spectrum, $f(p,t)\sim T^\ast p_\ast^2/p^3$ (as can be seen by expanding Eq.~(\ref{solutioninel}) for $p\gtrsim p_\ast$). This front is clearly visible in Fig.~\ref{fig:approximate1}: the successive times ($\tau_n=4\tau_{n-1}$) are chosen so that successive curves are equally spaced in the logarithmic momentum scale, as they should if $p_\ast\sim\sqrt{\tau}$. 

It is interesting to observe that, while the underlying microscopic mechanisms are different,  the picture that we get here for the thermalization of the soft region is not too different from that obtained earlier in the case of elastic collisions: thermalization proceeds from $p=0$, and the expansion of the soft, thermalized, region is controlled by a scale $p_\ast(\tau)\sim \sqrt{I_a \tau}$.
One sees on Fig.~\ref{fig:approximate1} that this soft region, with  $p\lesssim 0.3 $, is thermalized  for values of $\tau$ of order unity. Until this time, the density stays approximately constant. Again, as it was in the elastic case, this is natural: it is only when the soft region is thermalized that the system ``knows" whether it is over or underpopulated.

The variation of the density  with $\tau $ can be semi-quantitatively understood by integrating Eq.~(\ref{approxequation}) in a sphere of radius $p_0$, and using the approximate solution (\ref{solutioninel}), which ignores the variation of $T^*$ with time. Calling $n_{p_0}$ the number density of particles in the sphere $p_0$, one gets
\beq\label{np01}
\frac{\rmd n_{p_0}}{\rmd \tau}&=&\frac{I_a R}{2\pi^2}\int_0^{p_0}\rmd p\, \left( \frac{T^\ast}{p} - f_0\right)\rme^{-p_\ast^2/p^2}\nn
&\approx& \frac{I_a R}{2\pi^2}\int_{p_\ast}^{p_0}\left( \frac{T^\ast}{p} - f_0\right)\nn
&\approx& \frac{I_a R}{2\pi^2}\left\{ T^* \ln\frac{p_0}{p_\ast}-f_0(p_0-p_\ast)  \right\}.
\eeq
In the second line, we have used the fact that the exponential factor cuts off the integrand at $p\lesssim p_\ast$, with $p_\ast=\sqrt{ RI_a\tau}$. This formula shows that the rate of particle production is infinite at the initial time $\tau=0$, and decreases as $p_\ast$ increases, that is as $\tau$ increases. It then vanishes when $p_\ast$ crosses $p_0$: at this point the particles within the sphere $p_0$ have reached thermal equilibrium and their number remains constant if the temperature stays fixed. This crossing point occurs at a time $\tau_0$ given by $\sqrt{RI_a\tau_0}=p_0$.\footnote{ Note that the true integral giving $\rmd n_{p_0}/\rmd \tau$ (first line of Eq.~(\ref{np01})) does not actually vanish, but it becomes negligible when $p_\ast$ crosses $p_0$ at time  $\tau\approx\tau_0 $.}  
At  time $\tau_0$, the number density of particles that have appeared in the sphere $p_0$ is given by
\beq
\int_0^{\tau_0}\rmd\tau\frac{\rmd n_{p_0}}{\rmd \tau}
&\approx& \frac{I_a R}{2\pi^2}\left(  \frac{T^*}{2} -\frac{f_0}{3} p_0\right)\tau_0\nn
&\approx& \frac{p_0^2 f_0}{4\pi^2} \left(   \frac{T^*}{f_0} -\frac{2}{3} p_0\right).
\eeq
To proceed further, recall  that  $\bar p=T^*/f_0$, the momentum at which  the equilibrium distribution $T^*/\bar p$ equals the initial distribution $f_0$, characterizes the size of the soft region.\footnote{Note that in contrast to  Eq.~(\ref{baromega}), we define here $\bar p$ with the initial temperature, rather than with the equilibrium one. }
This scale exists only if $f_0$ is large enough so that $\bar p< Q_s$. Indeed, by using the relation (\ref{Tstar0}), $T^*=T^*_0=Q_s (1+f_0)/3$, we get $\bar p=Q_s\,(1+f_0)/(3f_0)$. In particular, for large $f_0$, $\bar p\simeq Q_s/3 <Q_s$. For small $f_0$ on the other hand, $\bar p\sim Q_s/(3f_0)$ which exceeds $Q_s$ when $f_0<1/3$. 

One can  then estimate the number of particles produced in the soft region by choosing $p_0=\bar p$ in the formula above. One gets
\beq\label{dnpodtauint}
\int_0^{\tau_0}\rmd\tau\frac{\rmd n_{\bar p}}{\rmd \tau}&\approx& \frac{f_0 \bar p^3}{12\pi^2}=n_{\rm in} \,\frac{f_0}{2} \left(\frac{\bar p}{Q_s} \right)^3.
\eeq
For large $f_0$, one has $\bar p/Q_s\simeq 1/3$, so that the density of produced particles represents a fraction $f_0/54$ of the initial density $n_{\rm in}$. In the opposite regime of small $f_0$, we have $\bar p/Q_s\simeq 1/(3f_0)$. However,  in this case, $\bar p$ is never reached, so that the fraction of produced particles is at most $f_0/2$, which is small since $f_0$ is small. Thus, in all cases, the number of  particles produced during the thermalization of the soft region is a small fraction of the initial density  (see also Fig.~\ref{fig:C-exact} below).  

\begin{figure}[!hbt]
\begin{center}
\includegraphics[width=0.75\textwidth]{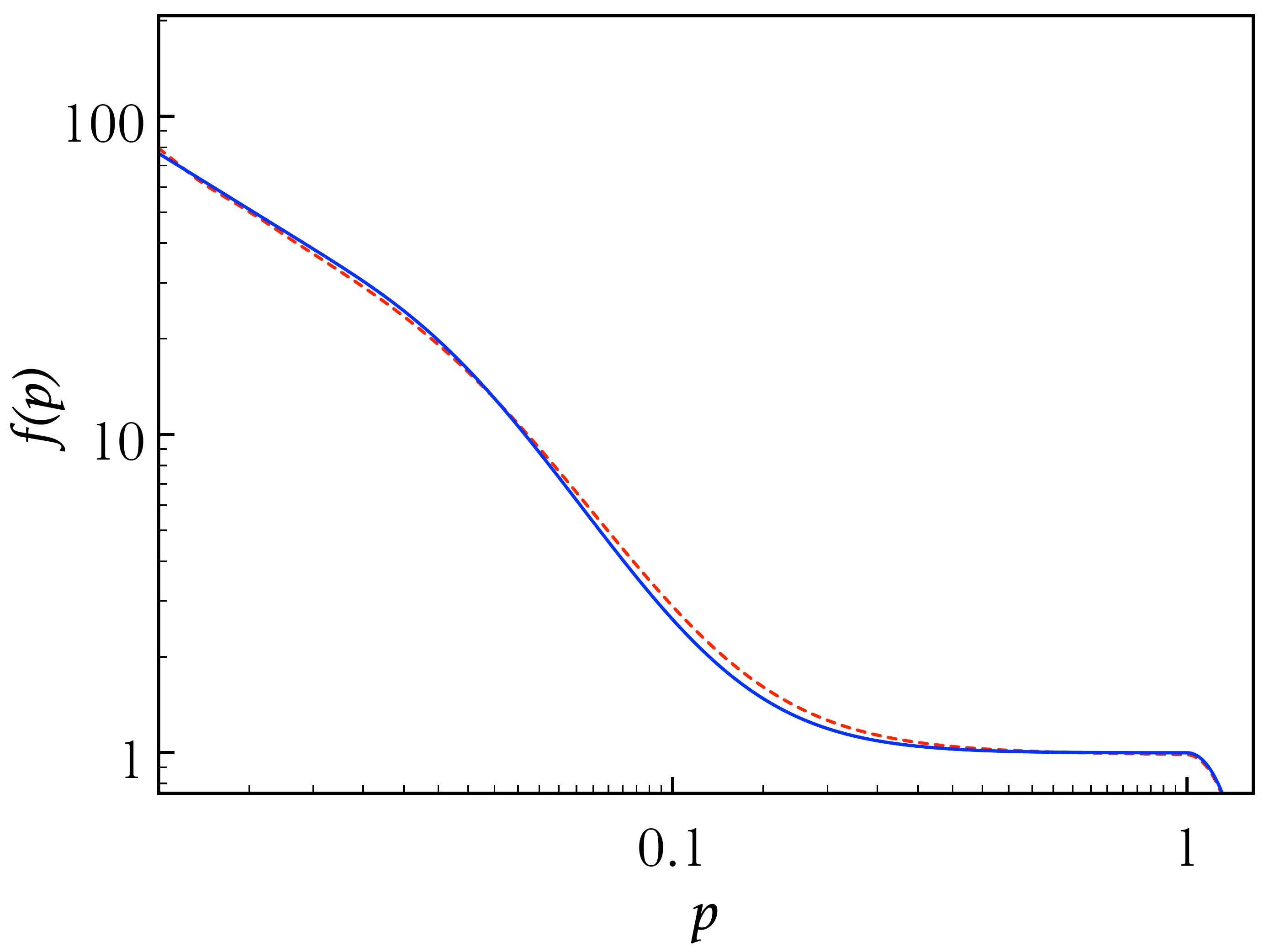}
\vspace{-0.1in}
\caption{(Color online). Comparison of the solution of the approximate equation (\ref{approxequation}) (blue curve) to that of the complete equation (dashed red curve), Eq.~(\ref{eq_D_BH_5}), at $\tau=0.04$.} \vspace{-0.25in}
\label{fig:comparison}
\end{center}
\end{figure}

One may extend this discussion to the case where the temperature $T^*$ is allowed to vary. We have
\beq\label{eq:Q-flux-inel0}
\frac{\rmd n_{p_0}}{\rmd \tau}=\frac{1}{2\pi^2} \int_0^{p_0} \rmd p\, p^2 \frac{\rmd f(p)}{\rmd\tau}=\frac{RI_a}{2\pi^2} \int_0^{p_0} \frac{\rmd p}{p} \left[T^\ast - p f(p)\right].
\eeq
When $p_0\lesssim p_\ast(\tau)$, the distribution is $f(p)\sim T^*/p$, so that the first equality yields
\beq
\frac{\rmd n_{p_0}}{\rmd \tau} \simeq \frac{p_0^2}{4\pi^2}\frac{ \rmd T^*}{\rmd\tau}.
\eeq
That is, when $p_0$ is smaller than $p_\ast$, the change of the particle number in the sphere $p_0$ is driven by the change in time of the temperature. 
In order to match with the second equality in Eq.~(\ref{eq:Q-flux-inel0}), we note that the last integral vanishes when $f(p)=T^*/p$. However there are corrections to the distribution function, which will be discussed more extensively in the next section. These corrections give a negligible contribution to the left hand side of Eq.~(\ref{eq:Q-flux-inel0}), but they contribute to the right hand side.  In particular, there is a correction linear in $p$ (as can be inferred from the general form (\ref{fullsolution}) of the solution). Assuming $f(p,\tau)\simeq T^*(\tau)/p+B(\tau) p$, and matching with the left hand side of Eq.~(\ref{eq:Q-flux-inel0}), one relates $B(\tau)$ to the rate of change of the temperature
\beq\label{BT*}
\frac{\rmd T^*}{\rmd \tau}=-R I_a B.
\eeq
Since $\rmd T^*/\rmd \tau<0$ (see Fig.~\ref{fig:IaIbTast} below), $B(\tau)>0$. 

\subsection{More results with the exact equation}

The previous discussion relies on the solution of the  approximate equation (\ref{approxequation}) which, we claimed, gives an accurate description of the small momentum region at small times, i.e., of the thermalization of the soft region. We now provide some numerical evidence supporting this claim, as well as further results that complete our description of the thermalization of the soft region. 

In Fig.~\ref{fig:comparison} the solution of the approximate equation (\ref{approxequation}) is compared to the solution of the complete equation, Eq.~(\ref{eq_D_BH_5}), obtained numerically. As can be seen the agreement is excellent, and this persists for all times $\tau\lesssim 1$. 

\begin{figure}[!hbt]
\begin{center}
\includegraphics[width=0.5\textwidth]{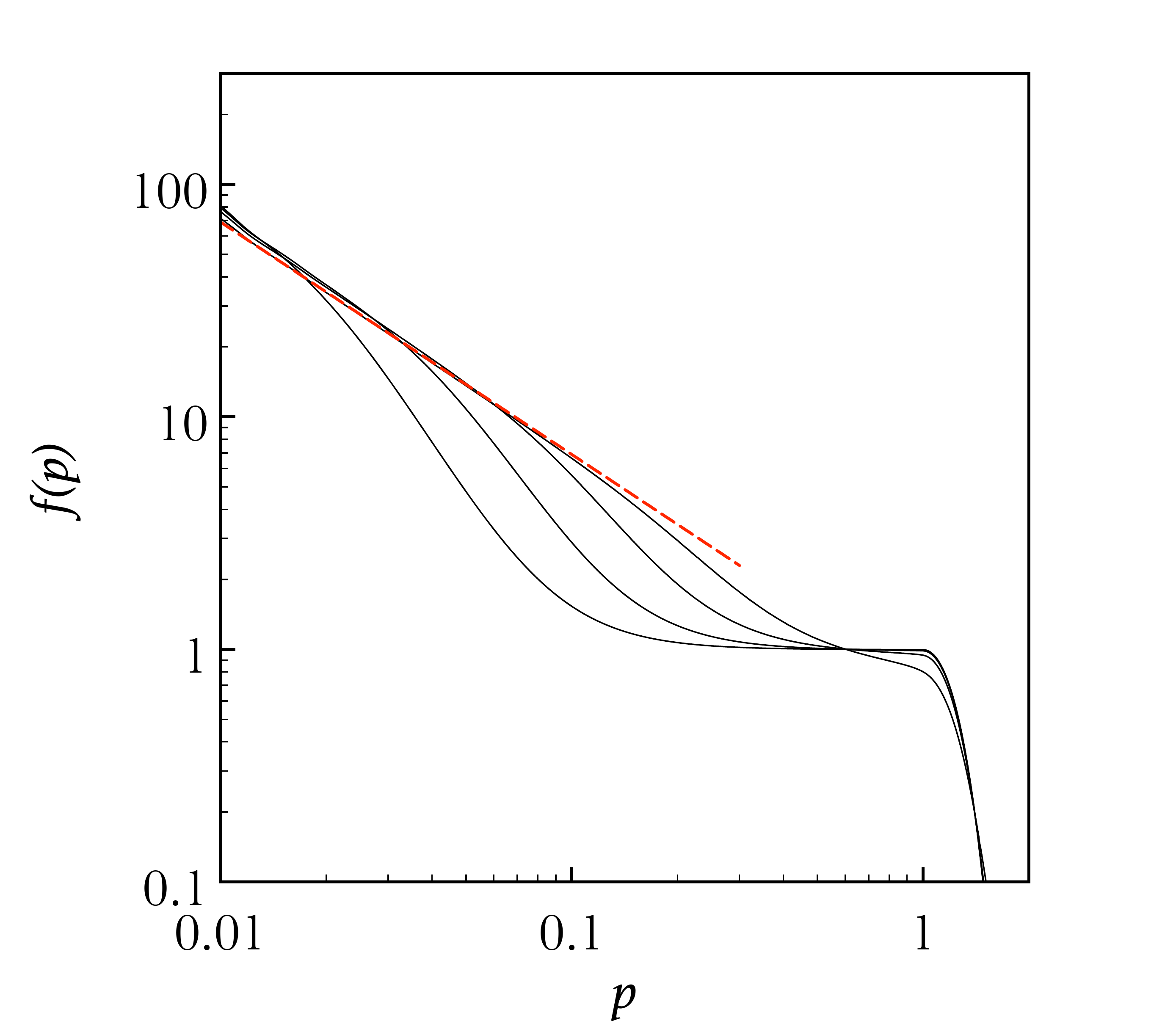}\includegraphics[width=0.5\textwidth]{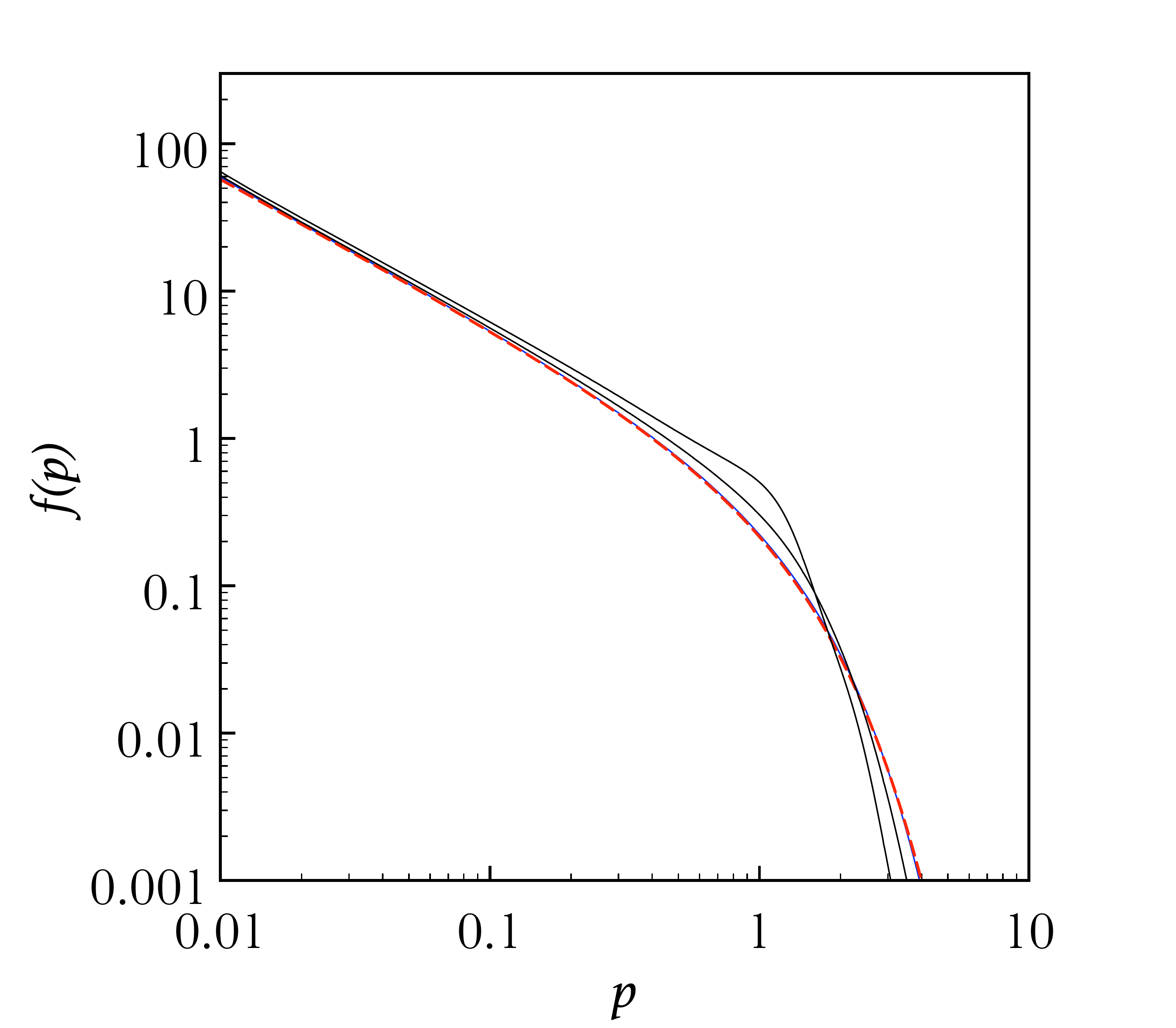}
 \vspace{0.0in}
\caption{(Color online). The solution of the exact equation  (\ref{eq_D_BH_5}) at early times.  Left panel: from bottom to top, $\tau=$ 0.01, 0.04, 0.16, 0.64. The  dashed (red) line   corresponds to to the thermal distribution $T^*/p$ with $T^\ast=0.69$ at $\tau=0.64$.  Right panel: from top to bottom, $\tau=$ 2.56, 8, 32. At the latest time $\tau=32.06$,  (blue curve), $f(p)$ is indistinguishable from the equilibrium Bose-Einstein distribution with $T_\text{eq}=T_\ast=0.58$  (dashed red curve).} \vspace{0.0in}
\label{fig:exactinelastic}
\end{center}
\end{figure}

\begin{figure}[!hbt]
\begin{center}
\includegraphics[width=0.5\textwidth]{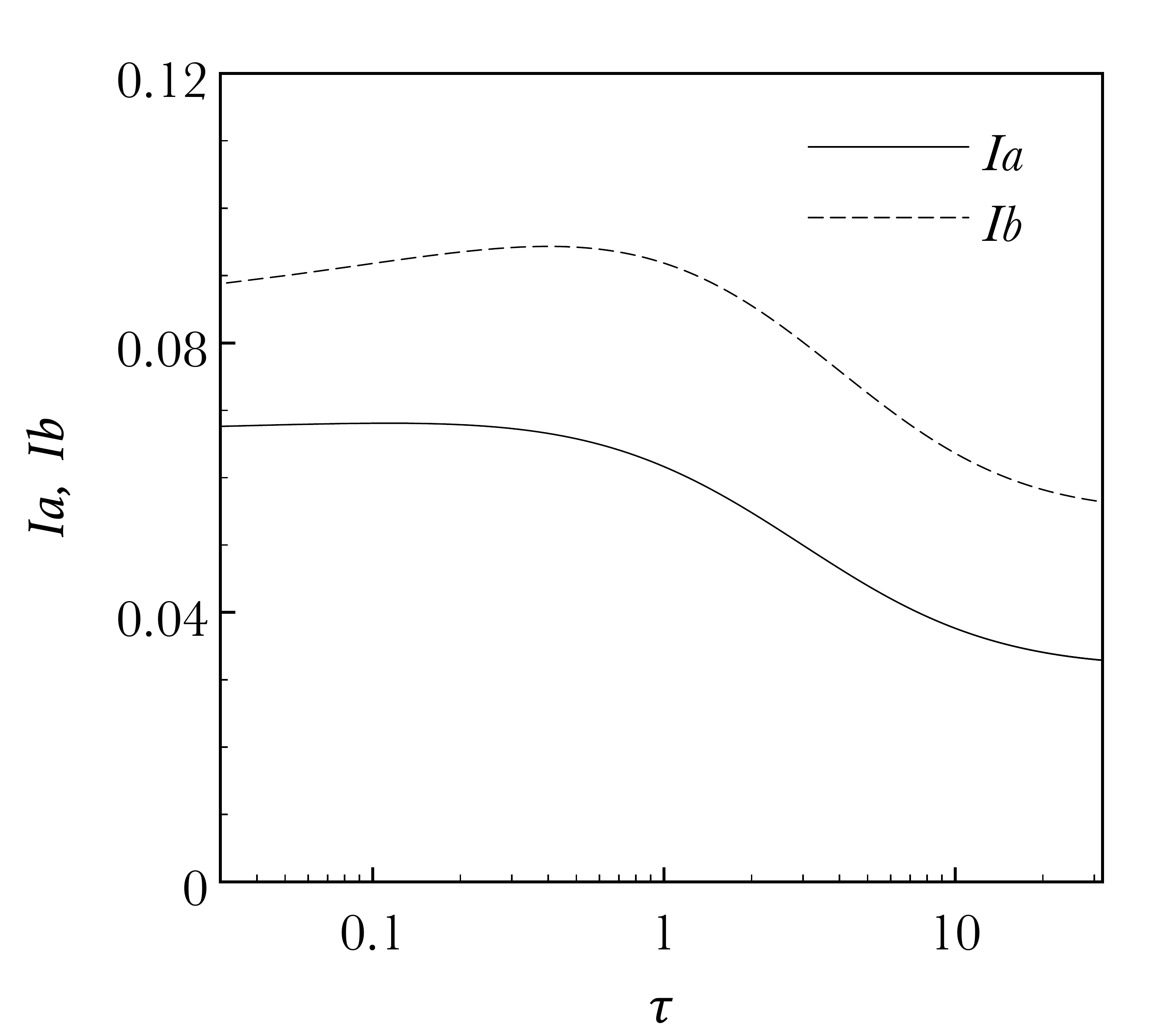}\includegraphics[width=0.5\textwidth]{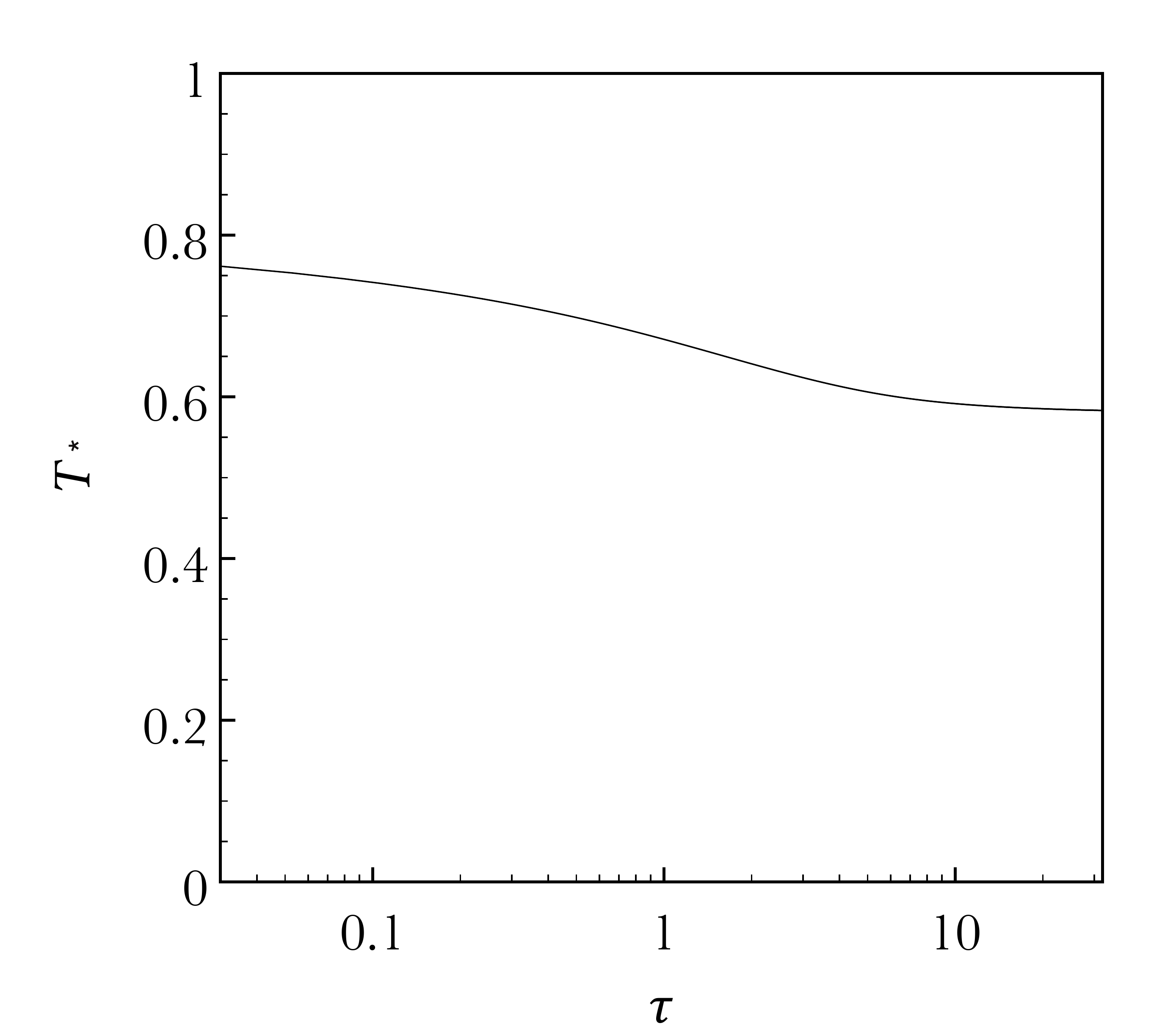}
 \vspace{-0.1in}
\caption{(Color online). The integrals $I_a$ and $I_b$ (left) and the effective temperature $T_\ast $ (right) as a function of $\tau$. The growth of the integral $I_b$ at small times can be attributed to the filling of the soft momentum modes to which $I_b$ is more sensitive than $I_a$. Note that the effective temperature of the soft region decreases slowly during thermalization, in contrast to what happens in the approximate equation.} \vspace{-0.25in}
\label{fig:IaIbTast}
\end{center}
\end{figure}

The solution at large time obtained from the exact equation is plotted in Fig.~\ref{fig:exactinelastic}. At short time, $\tau\lesssim 1$, the solutions of the exact and the approximate equations are nearly indistinguishable, as just said. However, in the exact solution,  the temperature drops regularly during the thermalization of the soft region (see Fig.~\ref{fig:IaIbTast} below), in contrast to what happens with the approximate equation where $T^*$ stays approximately constant when $\tau\lesssim 1$. At later times, the approximation that leads to Eq.~(\ref{approxequation}) become inaccurate. In particular, Eq.~(\ref{solutioninel}) does not conserve energy, and cannot therefore correctly describe the approach to equilibrium. Fig.~\ref{fig:exactinelastic} shows that, with the full equation,  the Bose-Einstein equilibrium distribution function is correctly reached at late time. 

\begin{figure}[!hbt]
\begin{center}
\includegraphics[width=0.5\textwidth]{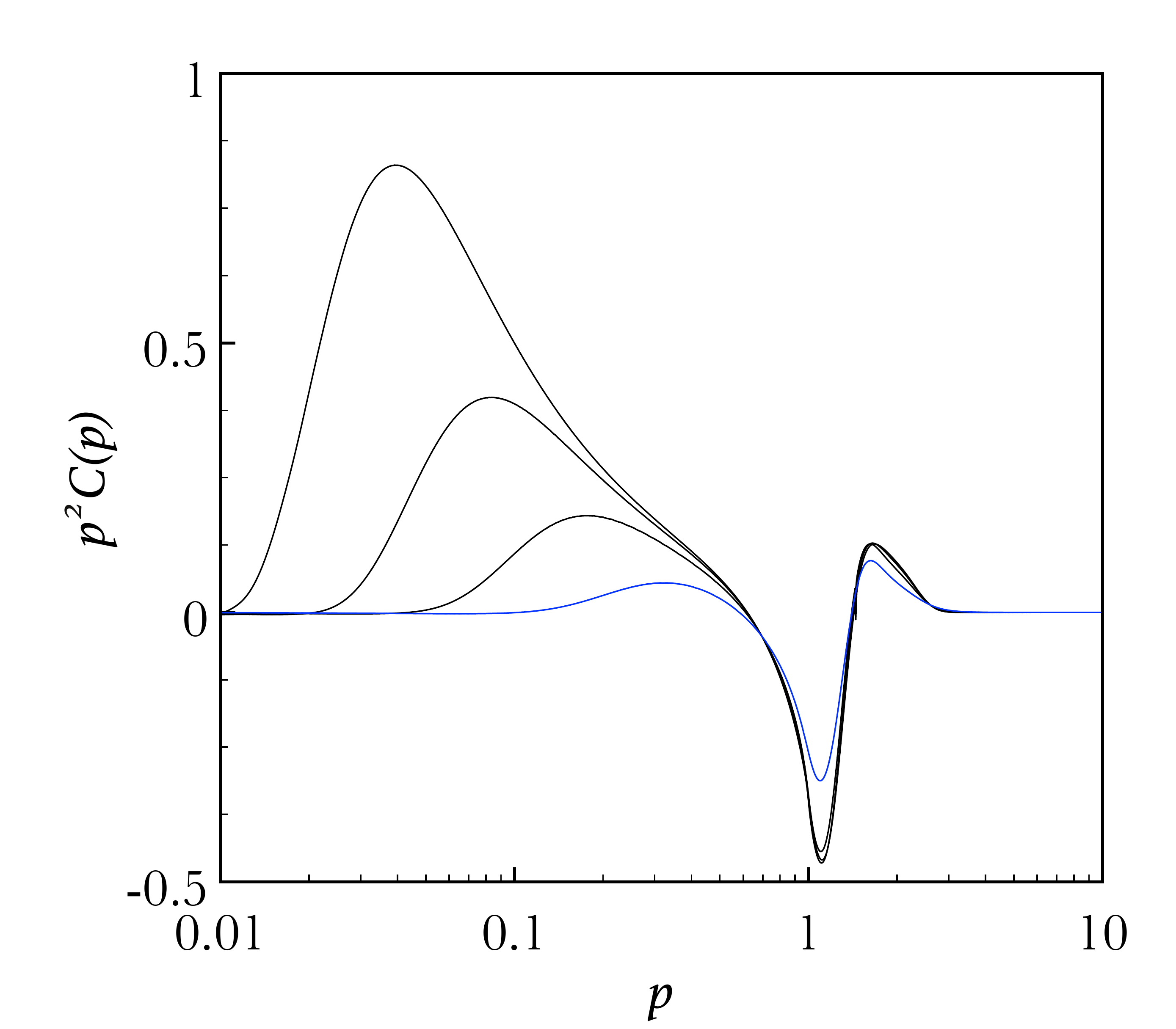}\includegraphics[width=0.5\textwidth]{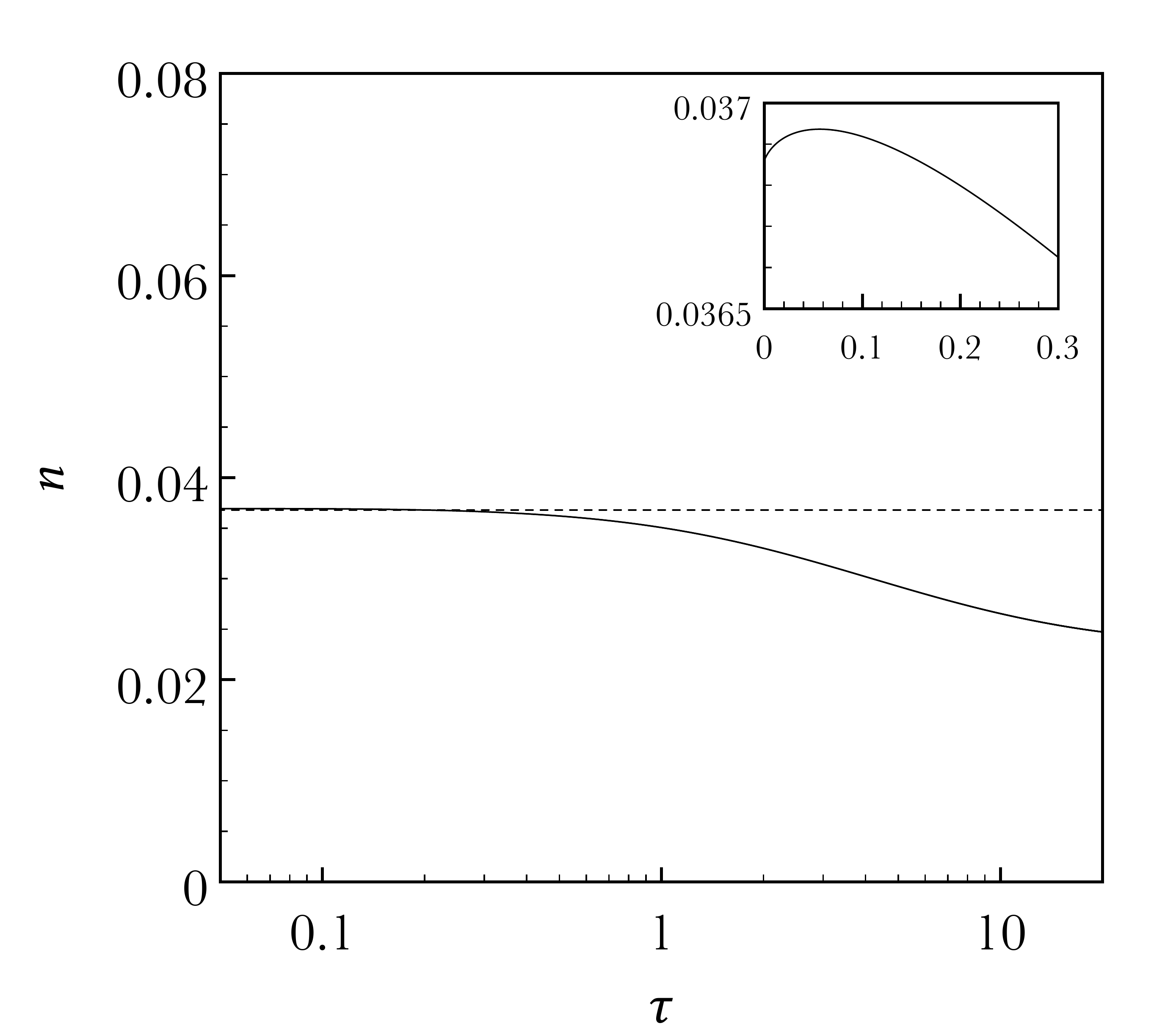}
 \vspace{-0.1in}
\caption{(Color online). The collision integral multiplied by $p^2$ and calculated with the exact equation for  $\tau=0.01, 0.04, 0.16, 0.64$ and $f_0=1$. On the right, the density of particles $n(\tau)$ is plotted  as a function of $\tau$.  } \vspace{-0.25in}
\label{fig:C-exact}
\end{center}
\end{figure}

Further insight can be gained by looking at the collision integral, multiplied by the phase space factor $p^2$. This directly measures the change in particle number locally in a spherical shell $[p,p+\rmd p]$ in momentum space. One sees clearly in  Fig.~\ref{fig:C-exact}, left panel, the growing region at small momentum where equilibrium is reached, and where consequently the collision term vanishes.  One sees also a large positive bump corresponding to the accumulation of soft particles radiated by the hard ones, hence the correlated dip around $Q_s$ associated with hard particles being slowly pushed towards smaller momentum as a result of their radiation of soft gluons.  As the soft particles are gradually eliminated, the collision integral decreases. The small bump at large momentum corresponds to particles that are pushed there by merging of two semi-hard particles, or the absorption of a soft particle by a hard one. 

The left panel of  Fig.~\ref{fig:C-exact} suggests that the total  number of particles increases  at short times (the integral under the curve measures the derivative with respect to $\tau$ of the total number of particles). The  inset plot in the right-hand-side figure shows that this is indeed the case, but this increase is very modest and invisible in the main plot. This is in line with our previous discussion based on the approximate equation (see the corresponding discussion around Eq.~(\ref{dnpodtauint})). As shown indeed in the right panel of Fig.~\ref{fig:C-exact}, the number of particles stays approximately constant during the thermalization of the soft region, and starts to decreases significantly only after the thermalization of the soft sector is completed, that is, when $\tau\gtrsim 1$.
Again, this is not unlike what happens in the purely elastic case, although there is here no sharp transition, like the BEC transition.

\section{Interplay of elastic and inelastic scattering}

By studying the thermalization driven  independently by elastic or inelastic processes, we have observed strong similarities, in spite of the very different processes involved. In both cases, thermalization proceeds first from the soft sector, starting instantly at $p=0$, and then developing into a region that spans all momenta up to a momentum of the order of 0.2 -- 0.3 $Q_s$. This soft momentum region is rapidly populated either by radiation from hard particles, in the inelastic case, or by the elastic drag current. Until the soft region is fully thermalized, and the system ``knows" about its overpopulation, the total number of particles stays approximately constant. After a time scale $\tau_c$ which is approximately the same in both the elastic and the inelastic cases, the excess particles start to be eliminated, via condensation in the purely elastic case,  via absorption (merging) of soft modes in the purely inelastic case.

There is however an important difference between the elastic and the inelastic cases: in the inelastic case, the distribution function is singular at $p=0$ at the start of the evolution, whereas in the elastic case, it takes the time $\tau_c$ for the effective chemical potential to vanish and for the distribution to become singular. Thus, when both processes are taken into account simultaneously, which is the case that we study in this section, one may expect the inelastic collisions to dominate at small $p$ and small $\tau$, and force $f(p)$ to be immediately singular. We shall verify shortly that this is indeed the case. 

In order to proceed with the small momentum analysis, we use first the approximate form of the inelastic collision term displayed in Eq.~(\ref{approxequation}). After multiplication by $p^2$, the  collision integral takes  then the form (at small enough $p$ to justify the replacement $f(1+f)\to f^2$ in the elastic current)
\beq\label{coll-term-full}
p^2 C[f]&=&p^2 C_\text{el}[f]+p^2 C_\text{inel}[f] \nn
&= &I_a  \frac{\del }{\del p} p^2 \left[\frac{\del  f(p)}{\del p} +\frac{f^2(p)}{T_\ast} \right]+ I_a R\left[\frac{T^\ast}{p} - f(p)\right].
\eeq 
A simple examination of this equation makes it clear that the presence of the elastic term does not change the fact that the solution must go to $T^\ast/p$ as $p\to 0$ in order  to avoid the blow up of the inelastic part of the collision integral. For this particular form of $f(p)$ the entire collision kernel actually vanishes, since for $f(p)=T^*/p$ the diffusion current exactly cancels the drag current. Note that this condition holds not only at early time, but for all times. As a result the flux of particles at the origin of momentum space always vanishes, indicating that condensation is hindered, however weak the strength of the inelastic processes may be \footnote{ In fact we shall argue later that the limit of weak inelastic collisions is singular.}.  Note that the condition $f(p)\sim 1/p$ is not enough to prevent condensation, in fact it is a prerequisite for condensation, and the fact that inelastic scatterings force this particular dependence on $p$ could even favor it (as was suggested in  \cite{Huang:2013lia}). The important point here is that the coefficient is $T^*$, and this is what, in addition to the functional dependence on momentum, is responsible for the vanishing of the current. In short, condensation is hindered because the system achieves equilibrium in essentially no time at $p=0$.

The important role of the inelastic processes  in forcing the form of the distribution function at the initial time reveals also important new features of the interplay of elastic and inelastic processes in the thermalization of the system. In the two cases considered previously, thermalization resulted from balance between reverse processes of the same nature, drag and diffusion for the elastic processes, splitting and merging for the inelastic ones. When both elastic and inelastic processes are present, the major competition appears to be between these two different processes, inelastic processes dominating initially. Note however that the major contributions to both the elastic and the inelastic collision integrals vanish, so that the competition will concern only the deviation from the thermal distribution $T^*/p$.

In order to describe how the soft momentum region of the system evolves in time, we look for a solution that deviates slightly from the fixed point solution $T^*/p$, and try an expansion of the form
\beq\label{ansatz-IR}
f(p,\tau)\simeq \frac{T_\ast(\tau)}{p}+ A(\tau) p^\alpha +B(\tau)p,
\eeq
where $\alpha $ is an arbitrary power, and $A$ and $B$ can be taken in general to be  analytic functions of $p$, but are restricted here to be simply functions of $\tau$.
The terms proportional to $A$ and $B$ are supposed to represent small corrections  to the leading contribution, in particular $Ap^\alpha$ should be less singular than $T^*/p$, i.e.,  $\alpha> -1$.  The term proportional to $B$ plays a special role; as shown above, it is related to the rate of change of the temperature $T^*$ (see Eq.~(\ref{BT*})).   

Using the ansatz (\ref{ansatz-IR}) in Eq.~(\ref{coll-term-full}), and keeping only the corrections linear in  $A$  and  $B$,  we find
\beq\label{coll-term-ansatz}
p^2 C[f]&\simeq & I_a A \left( \alpha^2+3 \alpha+2-R\right) p^\alpha+I_a B (6-R)p .
\eeq
The term linear in $p$ can be matched with a corresponding term in the left hand side of the kinetic equation, yielding  a slightly modified version of Eq.~(\ref{BT*}):
\beq
\frac{\rmd T^*}{\rmd \tau}=I_a B (6-R).
\eeq
Note that since $T^*$ is a decreasing function of time (see Fig.~\ref{fig:number-el-inel} below), $B(\tau)$ is now negative (for $R$ not too large), i.e., it has the opposite sign as in the purely inelastic case discussed earlier (Eq.~(\ref{BT*})). This is another illustration of the competing effects of elastic and inelastic scatterings.

Considering now the term in $p^\alpha$ in the right hand side of Eq.~(\ref{coll-term-ansatz}),  it is easily verified that there is no such contribution in $p^2\rmd f/\rmd\tau$ with which it could be matched. Therefore, for (\ref{ansatz-IR}) to be a solution, this term must vanish, which implies (with $\alpha>-1$) 
\beq\label{alpha-R}
\alpha = \frac{-3+\sqrt{1+4R}}{2}.
\eeq
For $R=1$, corresponding to comparable rates for elastic and inelastic processes, $\alpha\approx-0.38$.

\begin{figure}[!hbt]
\begin{center}
\includegraphics[width=0.5\textwidth]{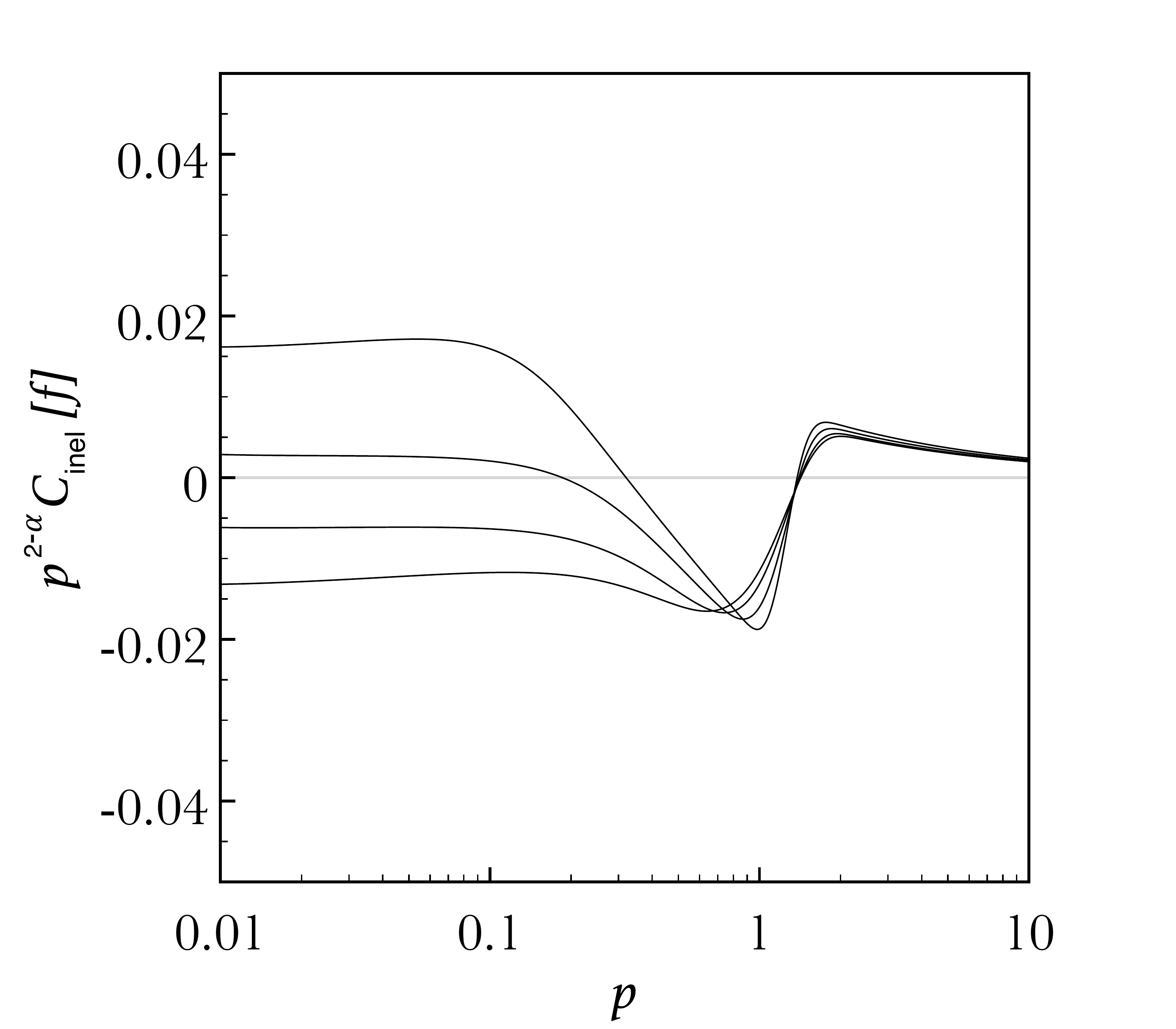}\includegraphics[width=0.5\textwidth]{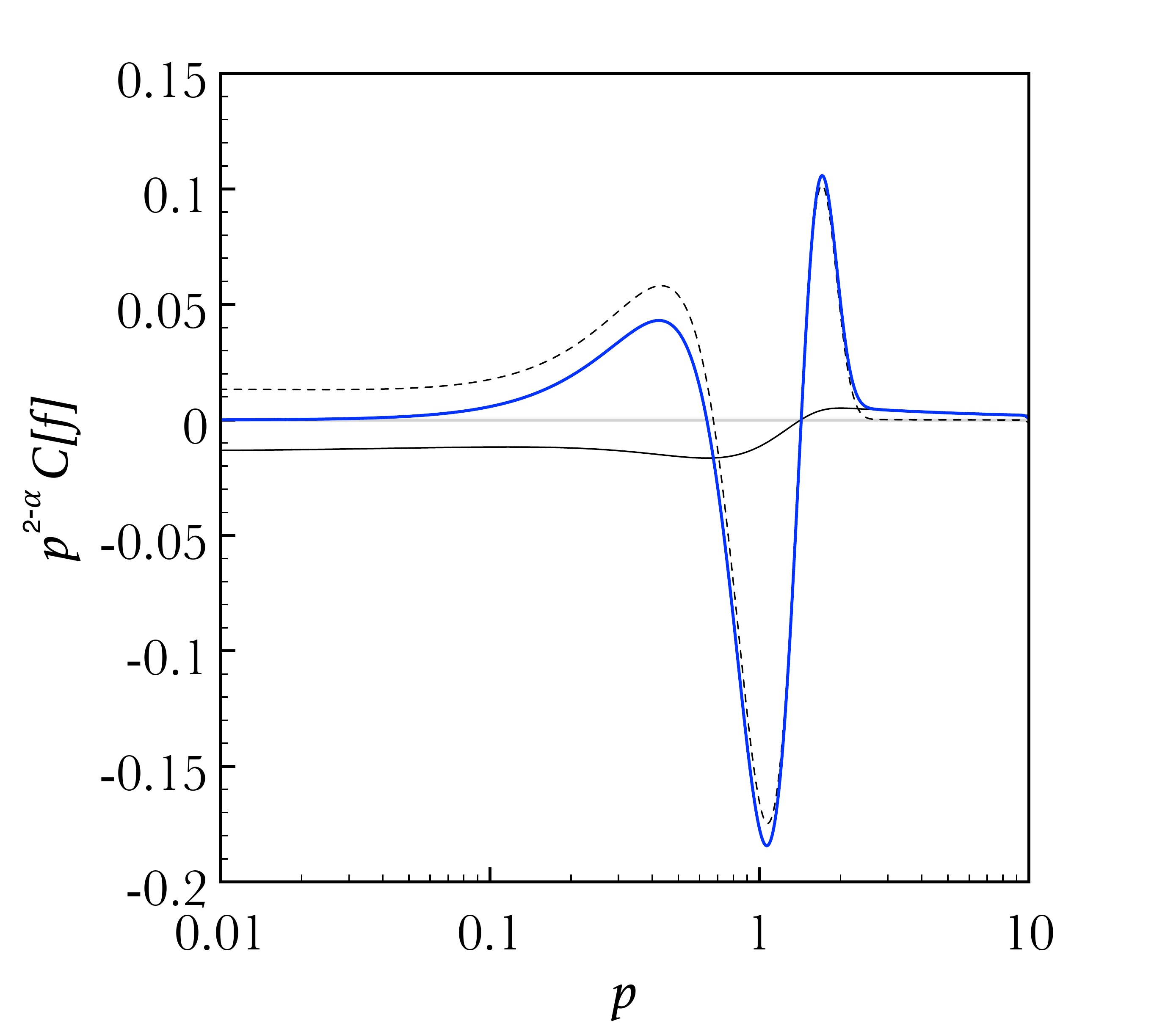}
 \vspace{-0.1in}
\caption{(Color online). Left panel: The approximate collision term (inelastic contribution) multiplied by $ p^{2-\alpha}$ with $\alpha\simeq -0.38$ (see text) at $\tau=$ 0.1, 0.4, 0.8, 1.1 (from top to bottom at $p=0.1$). The resulting curves are flat in the region of $p\sim 0.01-0.1$ in agreement with the inelastic contribution ($\sim AR$) in Eq.~(\ref{coll-term-ansatz}). In the right panel, the full and dashed curves correspond to the inelastic (negative)  and elastic (positive) contributions at $\tau=1.1$, that cancel each other in the infrared as shown  by the thick blue curve that depicts the full collision term.  }
\vspace{-0.25in}
\label{fig:el-inel-coll-model}
\end{center}
\end{figure}

This behavior is supported by the  numerical solution of the kinetic equation with the approximate collision term, Eq.~(\ref{coll-term-full}). In Fig.~\ref{fig:el-inel-coll-model}, left panel, the quantity  $p^{2-\alpha} C_\text{inel}[f]$ is plotted as a function of $p$. For small momenta, it is well approximated by the terms in Eq.~(\ref{coll-term-ansatz}) that are proportional to R, $p^{2-\alpha} C_\text{inel}[f]\approx -I_aAR-I_aBRp^{1-\alpha}$. In particular, the constant behavior $\sim -I_aAR$ (the term in $B$ is negligible when $\alpha<1$) is clearly visible  in the soft sector ($p \simeq 0.01,0.1$).  From this plot, we immediately deduce that $A(\tau)$ is initially negative and grows as a function of time. This behavior is natural. Indeed at short times, as we already argued, inelastic collisions dominate, and  populate the soft region, hence the positive collision term, and accordingly the negative signe of $A$. Note that, in this regime, the elastic current removes particles from the soft region, as necessary to balance the effect of inelastic collisions and drive the system to equilibrium. This is opposite to what we could have expected, given the form of the initial drag current. However, we should recall that the dominant contributions to the drag and diffusion current mutually cancel each other, and the remaining, leading  contributions take the form
\beq\label{currentA}
{\cal J}(p)=-I_a A(\alpha+2)p^{\alpha-1}.
\eeq
The term proportional to $\alpha$ is the diffusion contribution. The other contribution comes from the drag, and may be viewed as an ``interference" term between the leading contribution of the distribution, $T^*/p$, and the correction $Ap^\alpha$. As a result, the contribution to the drag adds to that of the diffusion, and both contribute to a positive current when $A<0$.

 As time passes the role of elastic collisions becomes eventually predominant, and the sign of $A$ turns positive. This occurs approximately for $\tau_c\gtrsim 0.4$. From that point on, the drag current pushes particles to the infrared, while the inelastic collisions eliminate them. The picture then is somewhat analogous to that of condensation, with the flow of particles through the origin of momentum space being replaced by a ``sink" extended over the entire soft region where   particles are eliminated by inelastic processes.   The thermalization of the soft region, indicated by the vanishing of the collision term, is illustrated in the right panel of Fig.~\ref{fig:el-inel-coll-model}. 
 
\begin{figure}[!hbt]
\begin{center}
\includegraphics[width=0.5\textwidth]{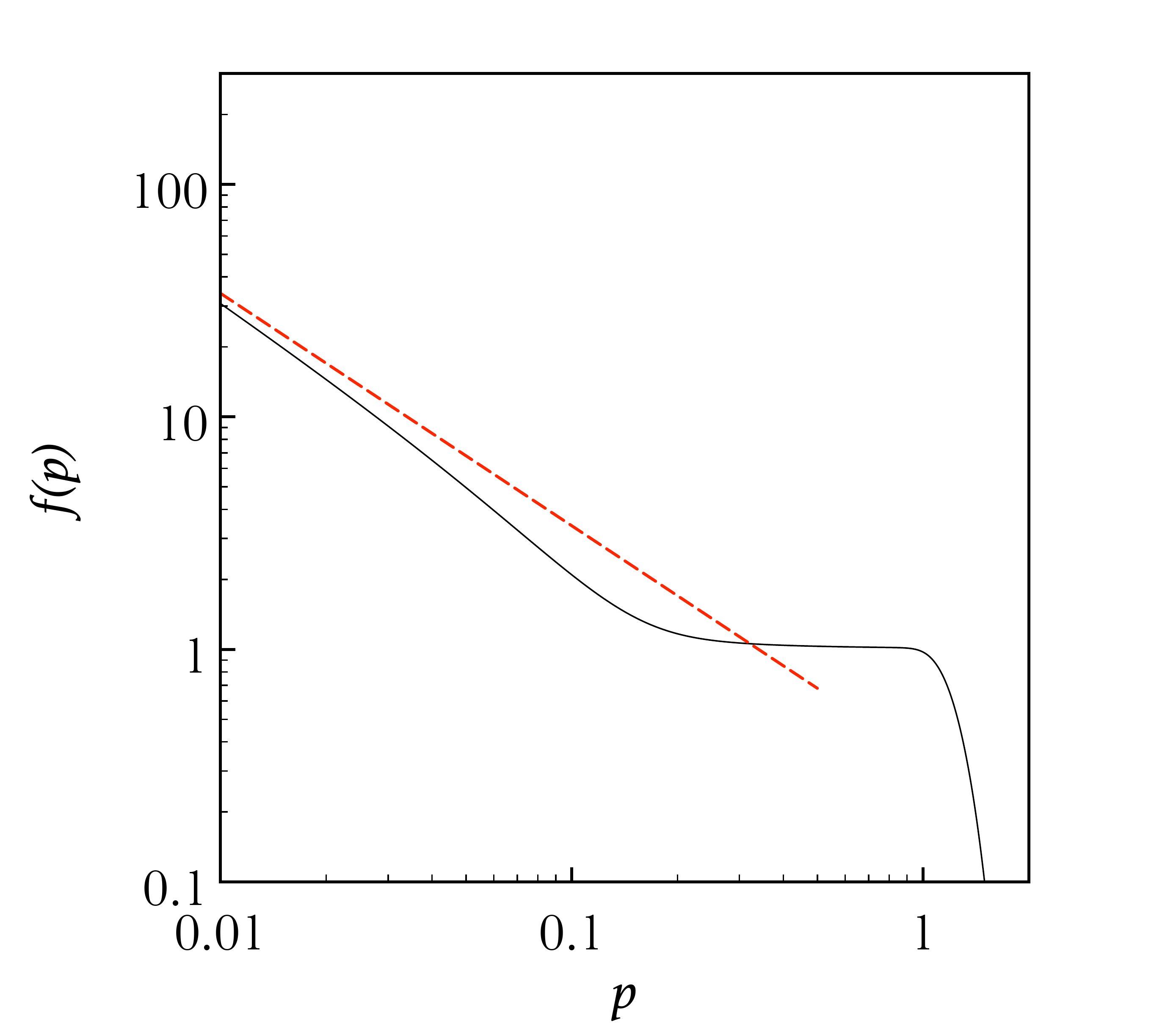}\includegraphics[width=0.5\textwidth]{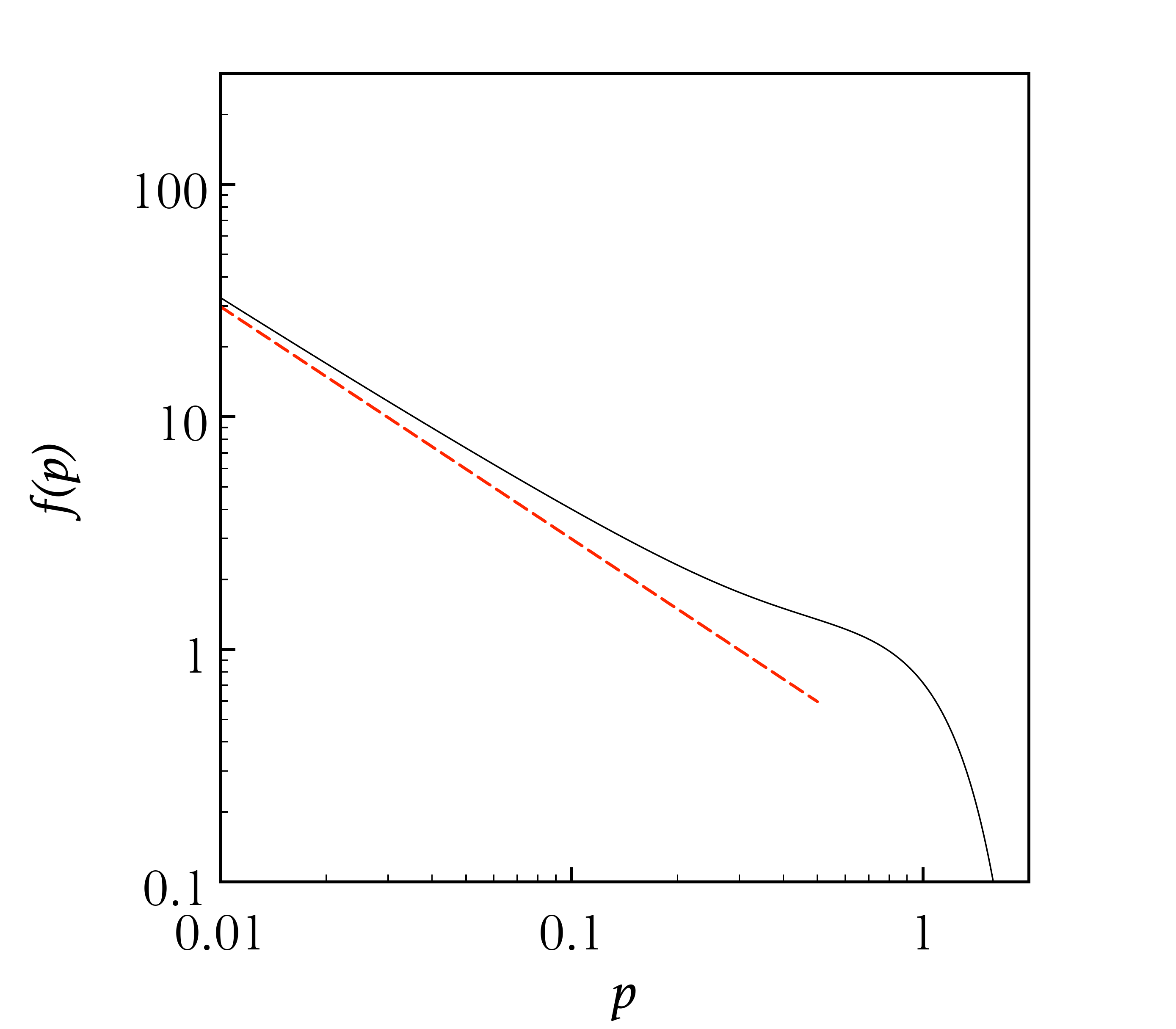}
 \vspace{-0.1in}
\caption{(Color online) The occupation number at $\tau=0.1$ (left) and $\tau=1.1$ (right), compared to the thermal spectrum with $T_\ast=0.34$ and $T_\ast=0.30$, respectively (dashed red lines). } \vspace{-0.25in}
\label{fig:el-inel-model}
\end{center}
\end{figure}

 The form of the distribution function at two times where $A$ is respectively negative ($\tau=0.1$) and positive ($\tau=1.1$)  is illustrated in Fig.~\ref{fig:el-inel-model}.  For $\tau=0.1$ (left panel) the equilibrium distribution is approached from below. This is the regime where $A<0$, corresponding to the population of the soft region by inelastic processes, while elastic collisions  have the opposite effect and pushes soft particles toward larger momenta. As a result the spectrum  tends asymptotically to the thermal limit from below.   At time $\tau=1.1$ the system is in the regime where $A>0$ dominated by elastic collisions. The correction to the equilibrium distribution is now positive. Note that this competition between inelastic and elastic processes, making the distribution smaller then larger than the thermal distribution,  appears to reduce the size of the soft region, which is limited here to smaller momenta ($p\lesssim 0.1$) than in the cases analyzed in the previous two sections.

Further insight into the competition between elastic and inelastic scattering encoded in the coefficient $A$ is gained by considering the density of particles $n_{p_0}$ in a small sphere of radius $p_0$ centered around $p=0$.  Leaving aside the contribution of the $B$ term, whose role has already been commented, we note that the current (\ref{currentA}) yields the following flux of particles on the sphere $p_0$
\beq
{\cal F}(p_0)=-\frac{I_a }{2\pi^2} A(\alpha+2)p_0^{\alpha+1}.
\eeq
 Since, as we have seen, these terms do not contribute to the rate of change of particles in the sphere $p_0$, this flux has to be compensated by the inelastic source, that is we must have, with ${\cal F}(p_0)$ given by the expression above, 
 \beq
{\cal F}(p_0)= - \frac{RI_a A}{2\pi^2}\frac{p_0^{\alpha+1}}{\alpha+1},
\eeq
which is indeed verified for  $\alpha$  given by Eq.~(\ref{alpha-R}).
When $A<0$ the inelastic source is positive: inelastic collisions move particles into the sphere $p_0$.  The elastic flux, also positive when $A<0$, has the opposite effect: it  removes particles from the sphere $p_0$. In the case $A>0$ the elastic flux is negative, elastic collisions push particles inside the sphere $p_0$, while the inelastic term plays the role of a sink.

\begin{figure}[!hbt]
\begin{center}
\includegraphics[width=0.5\textwidth]{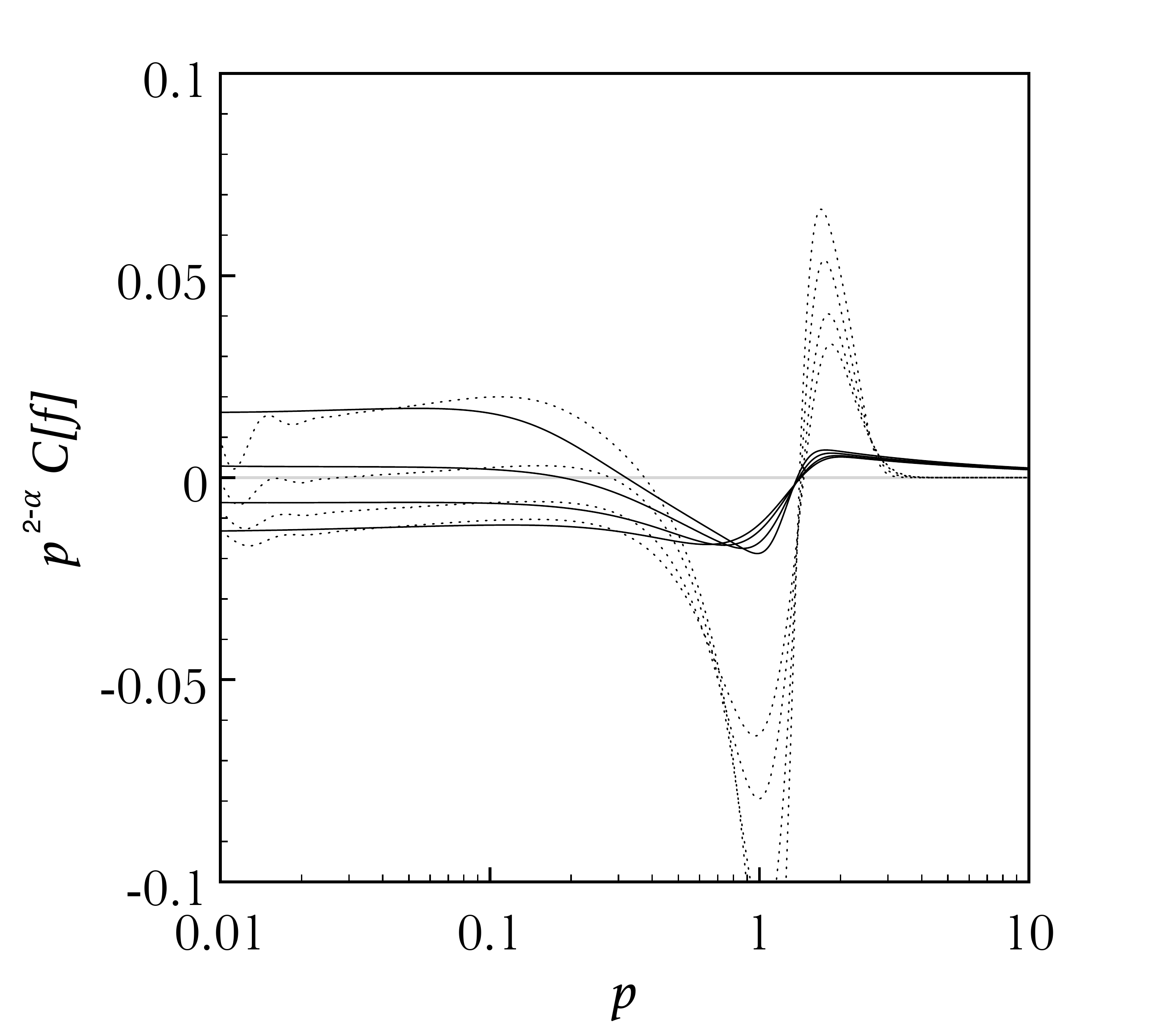}\includegraphics[width=0.5\textwidth]{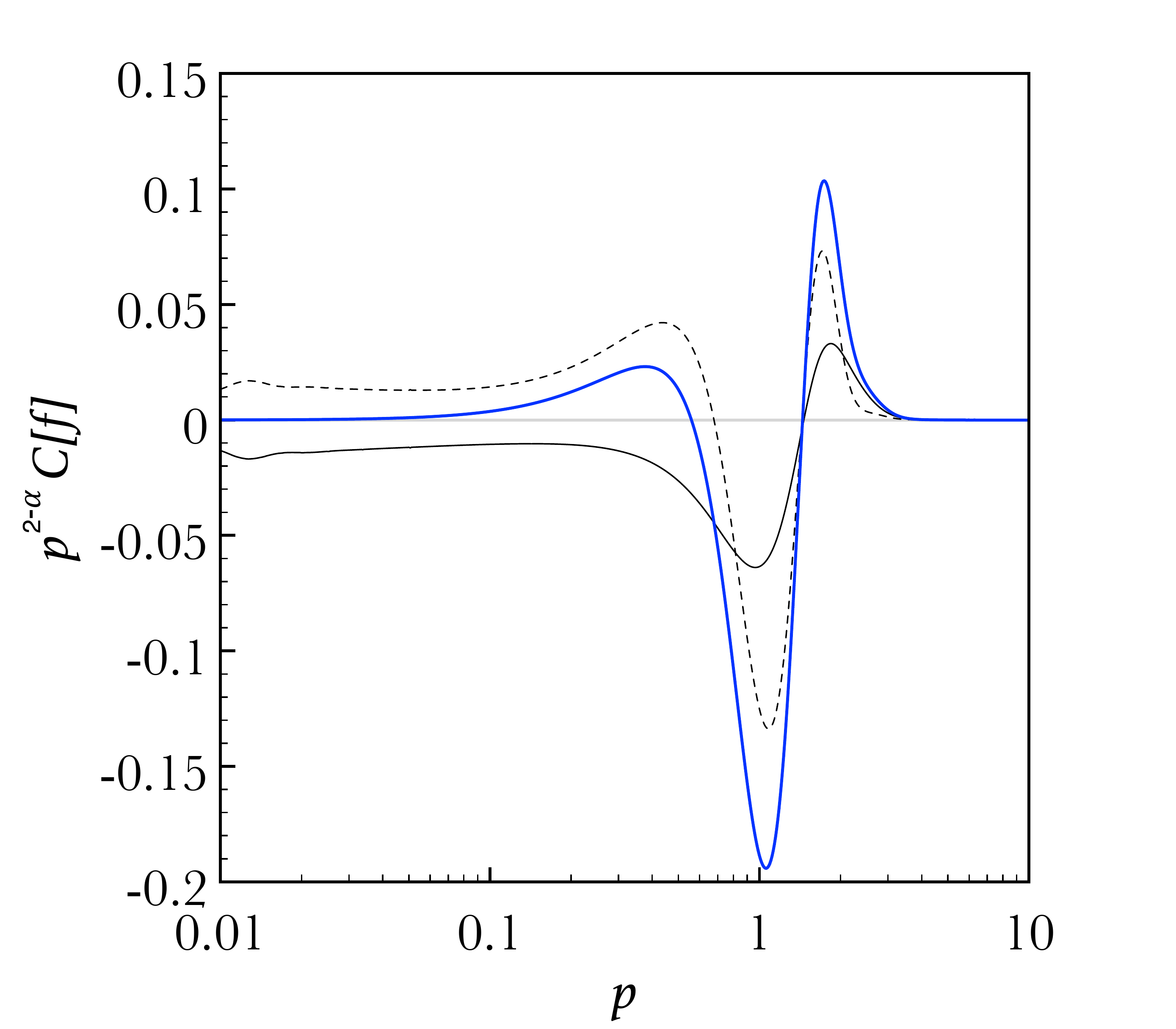}
 \vspace{-0.1in}
\caption{(Color online). Left panel: Comparing the exact inelastic collision term (dashed lines) to the approximation (\ref{coll-term-full}) at $\tau=$ 0.1, 0.4, 0.8, 1.1. The full lines correspond to the approximate solution (see left panel of Fig.~\ref{fig:el-inel-coll-model}).  Right panel: exact elastic and inelastic terms compared to the full collision term (same convention as in the right panel of Fig.~\ref{fig:el-inel-coll-model}). 
}
\vspace{-0.25in}
\label{fig:el-inel-coll-model/exact}
\end{center}
\end{figure}

\begin{figure}[!hbt]
\begin{center}
\includegraphics[width=0.5\textwidth]{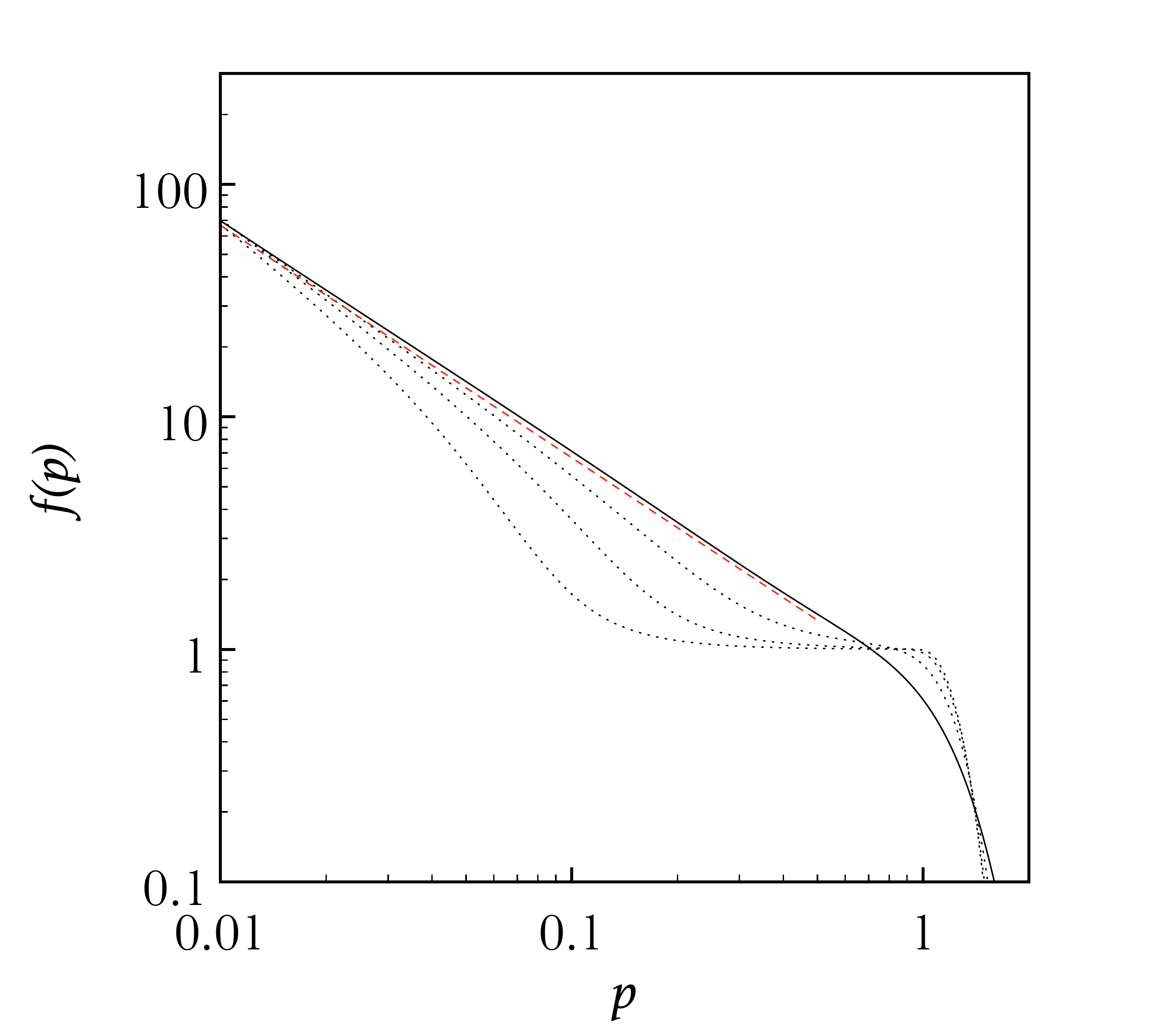}\includegraphics[width=0.5\textwidth]{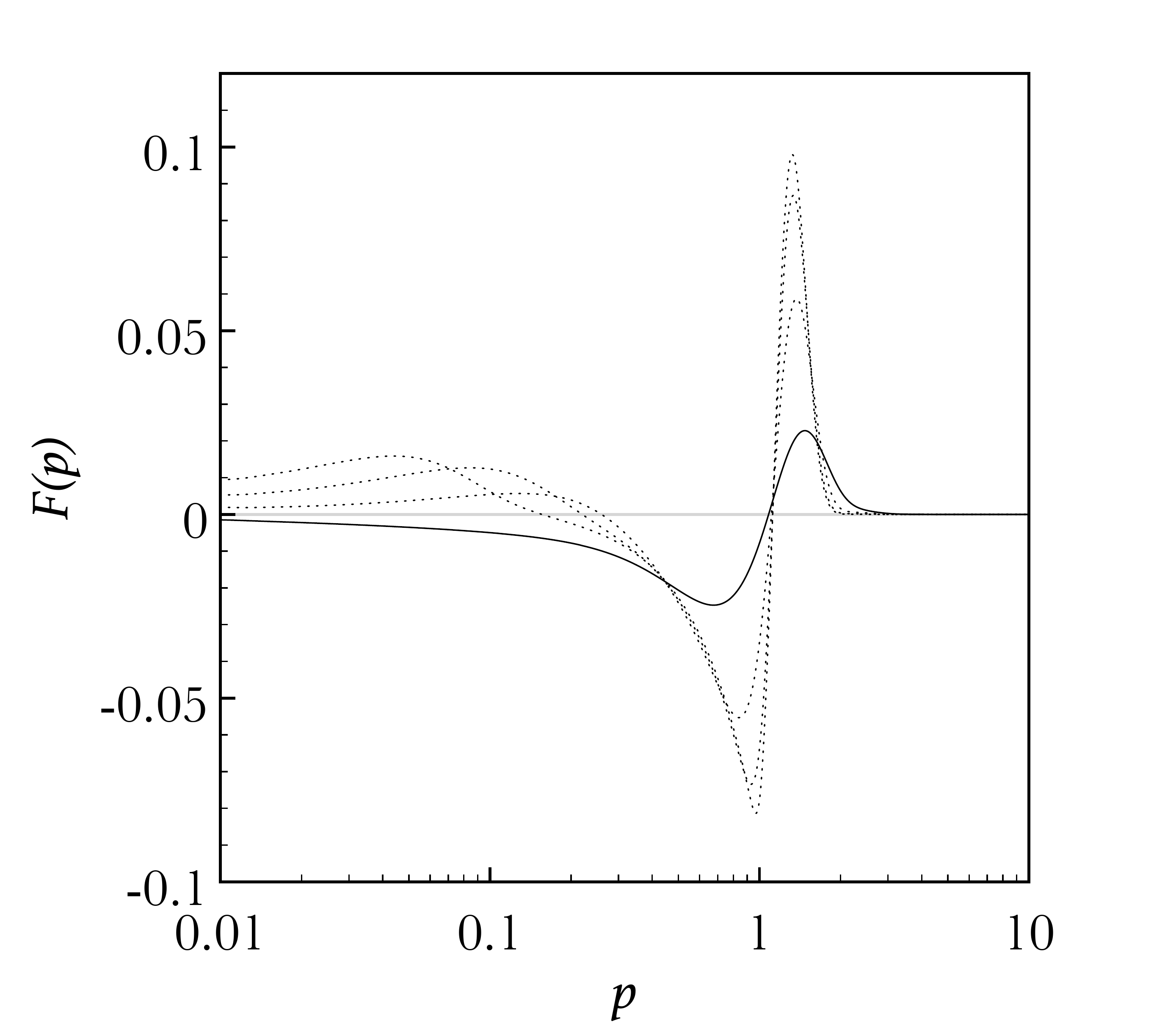}
 \vspace{-0.1in}
\caption{(Color online). The occupation number (left) and the elastic particle flux (right) at $\tau=0.01, 0.04, 0.16, 0.64$, from solving the exact kinetic equation, where the full curves correspond to the largest time $\tau=0.64$. The dashed red curve on the left panel stands for to the thermal distribution $T_\ast/p$ with $T_\ast \approx 0.68$ at $\tau=0.64$. The distribution $f(p)$ lies slightly above this curve, indicating that by the time $\tau=0.64$, the elastic collisions start to dominate and produce a negative flux, as confirmed by the right panel. } \vspace{-0.25in}
\label{fig:el-inel-occ}
\end{center}
\end{figure}

It is interesting to look at the limit of large and small $R$. When $R$ is large, e.g. $R>6$, then $\alpha>1$ and the correction to the distribution is negligible at small $p$. This limit appears to be smooth and corresponds to the situation where inelastic processes completely dominate. The approach to equilibrium of the soft sector ressembles that obtained in Sect.~\ref{sec:inel-scatt}, with minor corrections from elastic collisions. The limit $R\to 0$, on the other hand, is singular. For small $R$, we have from Eq.~(\ref{alpha-R}), $\alpha\simeq -1+R$, and the flux at $p_0$ takes the form
\beq\label{eq:Q-flux-inel3}
{\cal F}(p_0)=-\frac{I_a ARp_0^{R}}{2\pi^2}.
\eeq
It results from Eq.~(\ref{eq:Q-flux-inel3}) that the effect of the inelastic collisions never disappears, and the flux vanishes at $p=0$.  
Note however that for small $R$,   $ {\cal F}(p_0)$ is almost constant, and vanishes only near $p_0\to 0$. In this case, elastic collisions dominate, and if it were not for the singularity at $p_0=0$ which forces ${\cal F}(p_0)$ to vanish, a constant flux of particles would develop at $p=0$, as when the system undergoes condensation. Note that this analysis ignores the dependence of $A$ on $R$, in particular, it assumes that $A$ does not become singular as $R\to 0$ (we have numerical evidence that $A$ is a decreasing function of $R$). \\

\begin{figure}[!hbt]
\begin{center}
\includegraphics[width=0.5\textwidth]{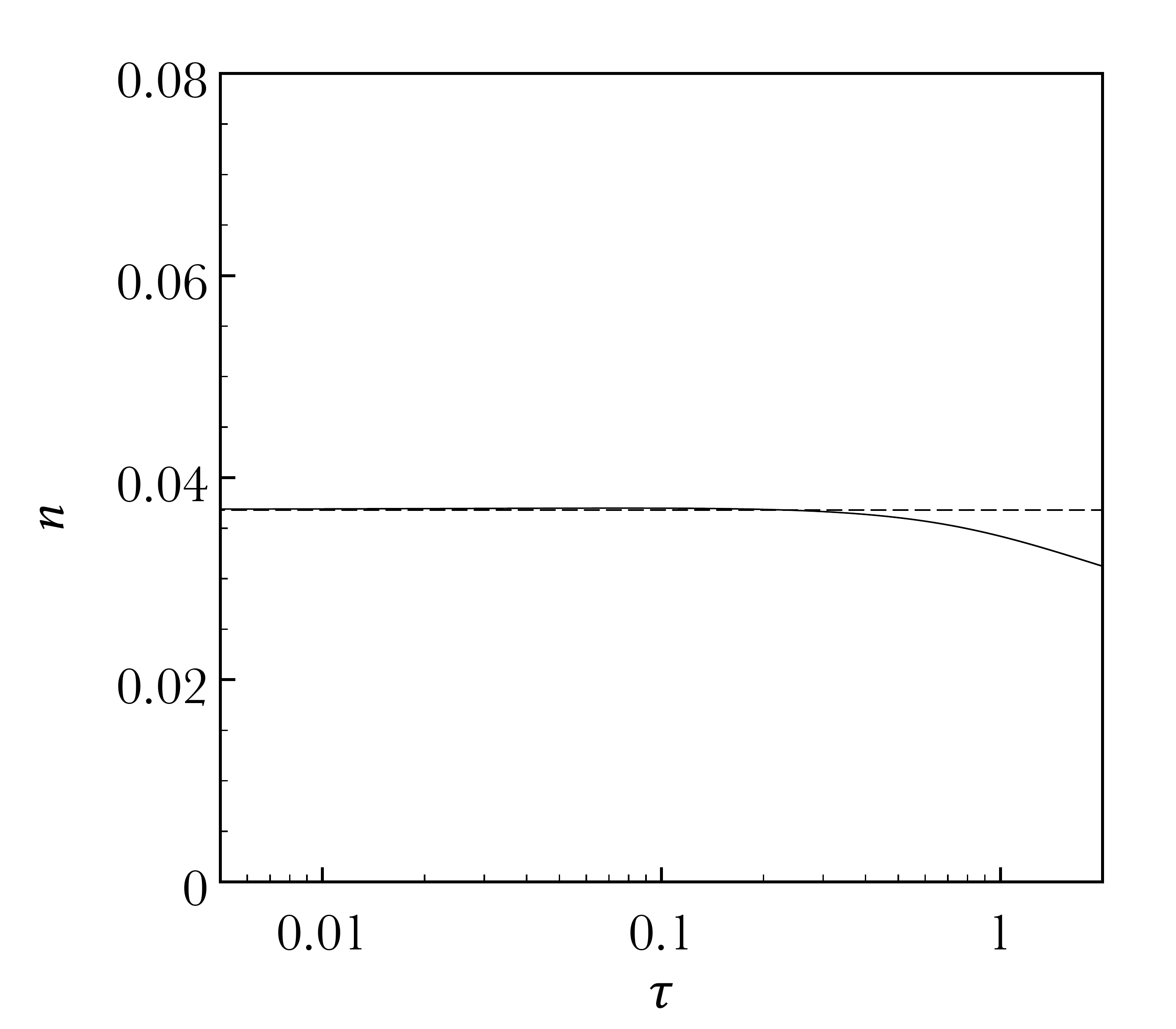}\includegraphics[width=0.5\textwidth]{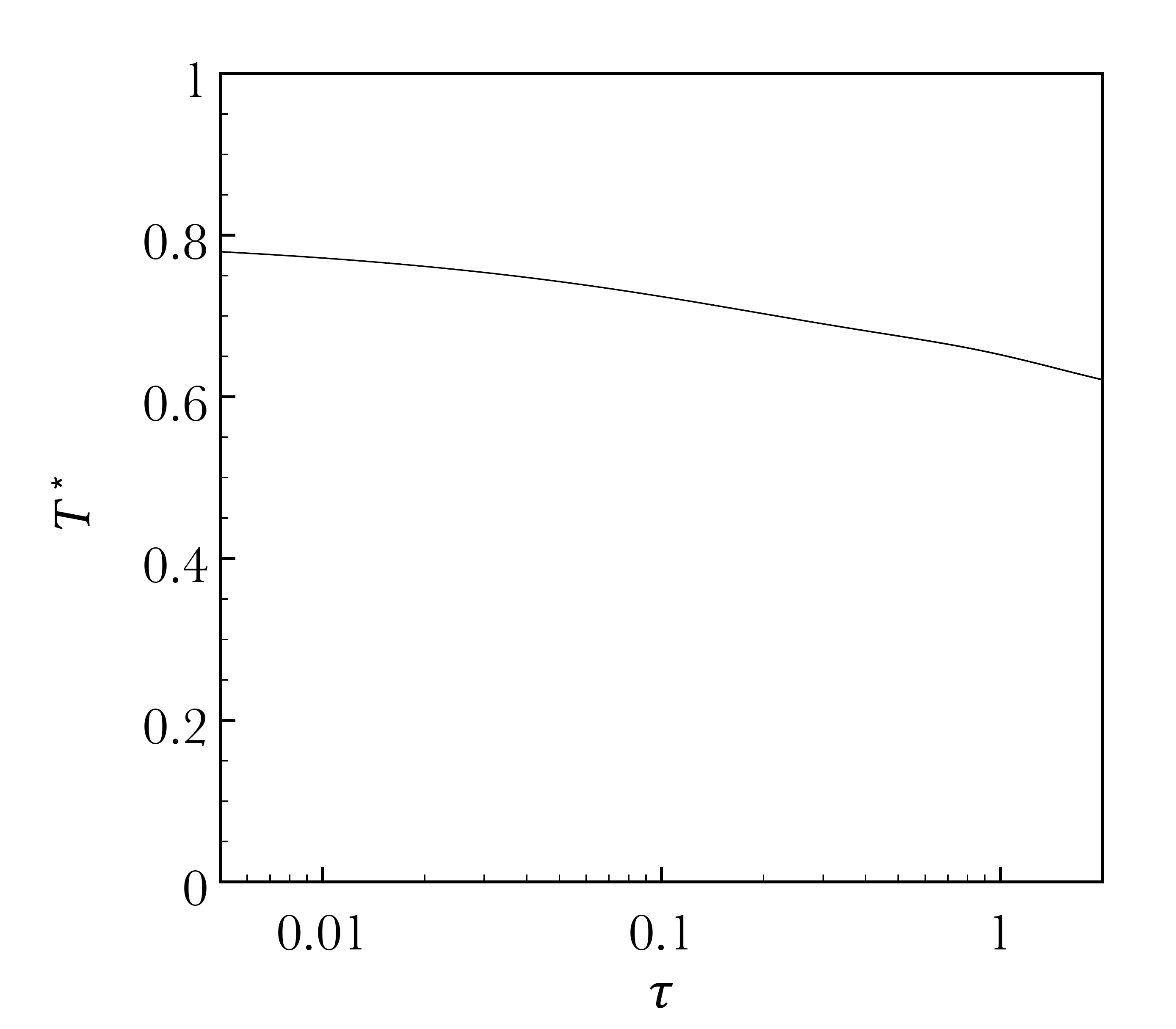}
 \vspace{-0.1in}
\caption{The density  of particles $n(\tau)$ (left) and the temperature $T^*(\tau)$ (right) as a function of time.  } \vspace{-0.25in}
\label{fig:number-el-inel}
\end{center}
\end{figure}

Until now, we have been using the approximate equation. We conclude this section by showing that the same behavior holds for the exact equation. This is illustrated in Figs.~\ref{fig:el-inel-coll-model/exact} and \ref{fig:el-inel-occ}.  Fig.~\ref{fig:el-inel-coll-model/exact} can be directly compared to Fig.~\ref{fig:el-inel-coll-model}.  It can be seen that the general pattern is well reproduced, and although the shapes of the functions are not identical, they are in qualitative agreement, and the time scales for the thermalization of the soft sector agree quantitatively. Finally Fig.~(\ref{fig:number-el-inel}) shows the variation with time of the particle density and the effective temperature $T^*$. Again we observe a delay for the density to start decreasing. This occurs sooner than in the previous two cases, at a time $\tau \simeq 0.4$ corresponding approximately to the time at which the coefficient $A$ changes sign.  By that time, the system seems to ``know" that it is overpopulated and particles start to be eliminated from the spectrum. Note however that the soft region is not quite equilibrated: the drag current continues to push particles towards the infrared, and the complete thermalization will occur at later time. To describe this phase fully, we need then to take into account more explicitly the hard particles.

 \section{Conclusion}
 
 In this paper, we have analyzed the thermalization of soft momentum modes in an isotropic, uniform and non expanding quark-gluon plasma. To do so, we have used approximate kinetic equations that we believe capture  essential features of QCD interactions: small angle elastic scattering, and inelastic number changing processes involving collinear radiation of  soft gluons.  The analysis reveals that thermalization occurs essentially instantly with the softer modes at $p\simeq 0$, and then gradually diffuse towards higher momentum modes. In both the cases where only elastic or inelastic processes are included in the kinetic equation, we observe a time delay before particles start to be eliminated. This delay we interpret as the time it takes to saturate the soft region which covers approximately all momenta up to $\sim 0.2-0.3 \,Q_s$. It is only when the soft region is equilibrated that the system starts eliminating particles. The same features are observed when both types of processes are included together, with however a shorter time scale. 
 
 In the competition between elastic and inelastic processes, which plays a dominant role when both processes are taken into account, it appears that the inelastic processes win at small times and small momenta. The main  reason is not so much that they are (parametrically) of the same order of magnitude as the elastic ones, but rather that the soft emissions induce very fast equilibration at small momenta. This is strong enough, within the present analysis, to block any flux of particles at $p=0$.

 The present work is limited to the soft modes, and we provided results for a limited range of values of the important parameters. We believe that for what concerns the soft modes, the conclusions that we have drawn are robust. However it is certainly of interest to understand better how things vary as we change the typical occupation $f_0$, the shape of the initial density profile, or the strength of the inelastic interactions. Also, the full description of thermalization implies the hard particles, i.e. particles with momenta of order $Q_s$.  
  Finally, to make a more direct contact with the phenomenology of heavy ion collisions,  the case of a longitudinally expanding system needs to be considered.  All these aspects can be handled within the same framework as that presented in this paper, and, together with a more detailed comparison with other related works, will be the subject of forthcoming publications.   
 
 \section*{Acknowledgements}
 The research of JPB is supported by the European Research Council under the Advanced Investigator Grant ERC-AD-267258. The research of JL is supported in part by the U.S. National Science Foundation under Grant No. PHY-1352368 and by the RIKEN BNL Research Center.  The research of YMT is supported by the U.S. Department of Energy under Contract No. DE-FG02-00ER41132.

\appendix
\section{Derivation of the kinetic equation}\label{derivationkinetic}

As indicated in the main text the types of inelastic processes that we consider in this work are dominated by nearly collinear emissions of soft gluons. These processes are quite analogous to those involved in the cascades of gluons that occur in  jets, and indeed the equations that we shall use to describe them are simple generalizations of the equations that  have been used in several works on jet physics (see \cite{Mehtar-Tani:2013pia,Blaizot:2015jea} and references therein). In this appendix, we show how one can easily deduce the kinetic equations from the equations that govern the inclusive distribution of gluons in a cascade. We shall consider two cases. 
We consider first the case where the branching time is small  compared to the typical elastic mean free path. Under this condition, a single scattering contribute to the radiation. We refer to this regime as the Bethe-Heitler (BH) regime. The second case concerns the case where multiple scatterings interfere in producing the final radiation. We refer to this as the Landau-Pomeranchuk-Migdal (LPM) regime.

We consider first the BH regime. Let us call $D(x,\tau) \rmd x$ the energy carried by gluons with energy $x=\omega/E$, with $E$ the energy of the gluon which initiates the cascade. We have \cite{Blaizot:2015jea} 
\beq\label{eqforDBH}
\frac{\del D(x)}{\del t}&=&C \int_0^1 \rmd z \frac{1}{z(1-z)}\left[ D\left( \frac{x}{z}    \right) -zD(x) \right]\nn
&=&C\int_0^1 \rmd z \frac{1}{z(1-z)}\left[ D\left( \frac{\omega}{z E}    \right) -zD\left( \frac{\omega}{E}\right) \right],
\eeq
where $C$ is  a constant that will be specified later. 
\begin{figure}[!hbt]
\begin{center}
\includegraphics[width=0.8\textwidth]{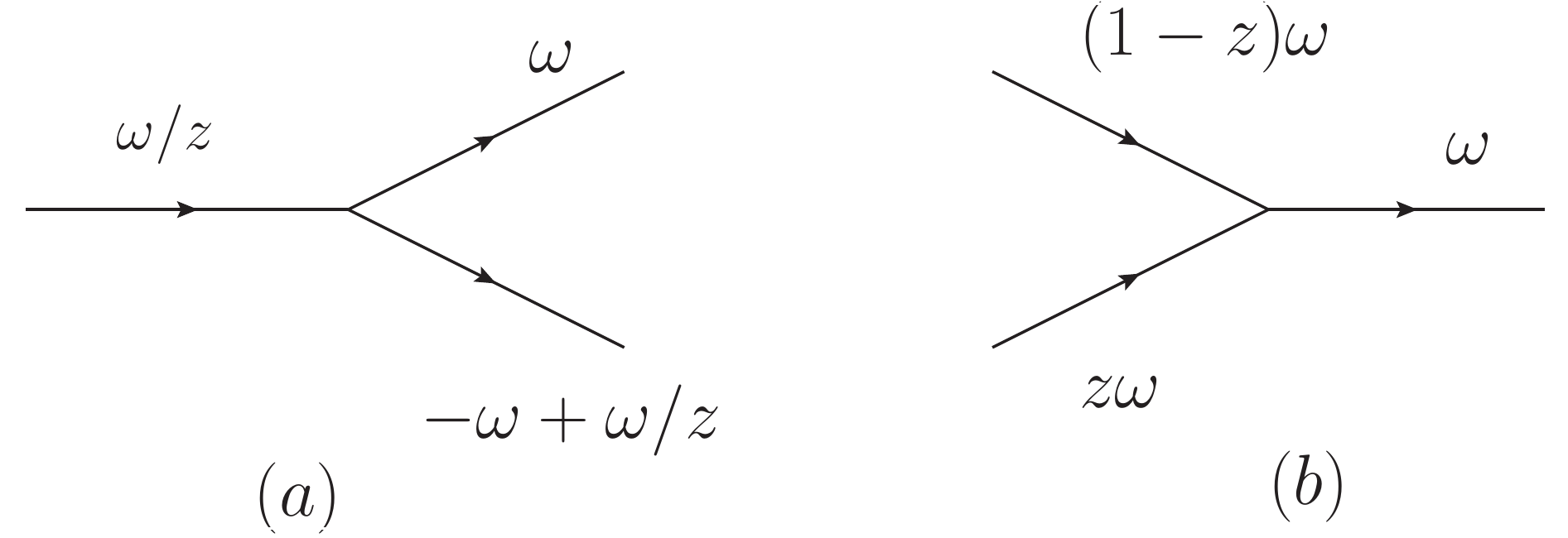}
\caption{Inelastic processes corresponding to gain terms (processes that populate the state with energy $\omega$) of splitting (left) and merging (right). The reverse processes corresponding to loss terms are not drawn.  } \vspace{-0.25in}
\label{fig:fig2}
\end{center}
\end{figure}
We then consider $D(\omega/zE)$ as a function of $\omega/z$ (since $E$ remains constant in all the discussion), and set
\beq
\omega^3 f(\omega) =\omega \frac{\rmd N}{\rmd \omega}= D\left( \frac{\omega}{E}\right),
\eeq
where $f(\omega) $ is the distribution function.
The equation above then becomes
\beq
\frac{\del f(\omega)}{\del t}=\frac{C}{\omega^3}\int_0^1 \rmd z \frac{\omega^3}{z(1-z)} \left[ \frac{1}{z^3}f\left( \frac{\omega}{z }\right) -zf(\omega) \right].
\eeq
In the first term ($\omega/z\rightarrow (\omega, \omega/z-\omega)$), we set $\omega'=\omega/z$, and get
\beq
\int_0^1 \rmd z \frac{1}{z(1-z)} \frac{\omega^3}{z^3}f\left( \frac{\omega}{z }\right)=\int_\omega^\infty \rmd\omega' \frac{\omega'^3}{\omega'-\omega} f(\omega')=\int_\omega^\infty \rmd\omega' K(\omega,\omega') f(\omega'),
\eeq
with 
\beq
K(\omega,\omega')\equiv \frac{\omega'^3}{\omega'-\omega} .
\eeq
In the second term (loss term, $\omega\rightarrow (z\omega, (1-z)\omega)$), we set $z\omega=\omega'$, and get
\beq
-\int_0^1 \rmd z \frac{1}{z(1-z)}z\omega^3f(\omega)=-\int_0^\omega \rmd \omega' K(\omega',\omega) f(\omega)
\eeq
The kinetic equation reads then
\beq
\frac{\partial f}{\partial t}=\frac{C}{\omega^3} \left\{ \int_\omega^\infty \rmd\omega' K(\omega,\omega') f(\omega')- \int_0^\omega \rmd \omega' K(\omega',\omega) f(\omega) \right\}.
\eeq
This equation describes the evolution of the distribution of gluons in a cascade of successive splittings. The first term in this equation is a gain term, with the splitting leading to an increase of $f(\omega)$. The second term is a loss term associated to a splitting that depletes $f(\omega)$.

\begin{figure}[!hbt]
\begin{center}
\includegraphics[width=1\textwidth]{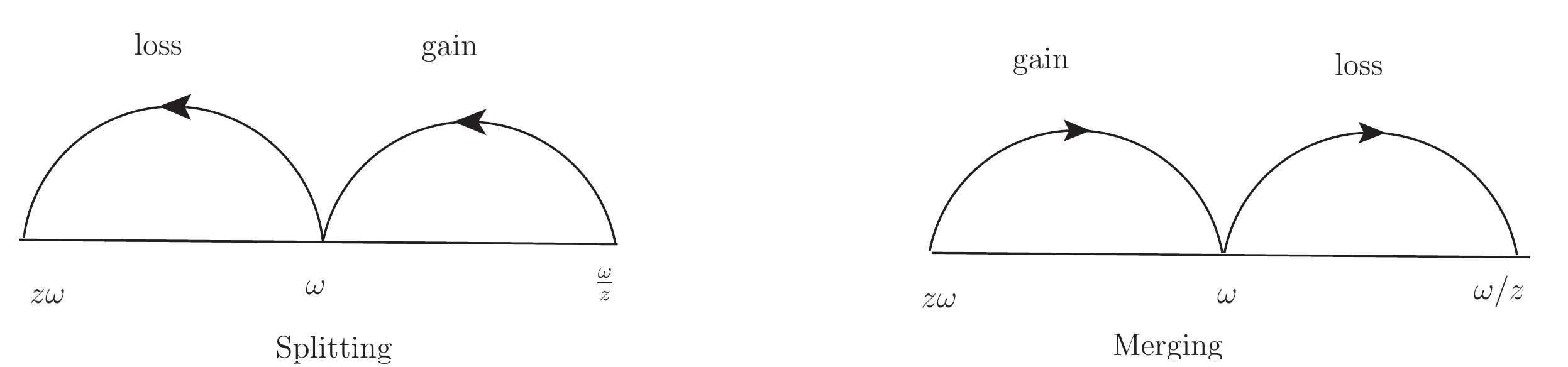}
\caption{Gain and loss terms for splitting and merging processes } \vspace{-0.25in}
\label{fig:gainloss}
\end{center}
\end{figure}
This kinetic equation is easily extended in order to treat properly the Bose statistics. This requires adding proper statistical factors, as well as including the merging processes (which involve both gain and loss terms -- see Fig.~\ref{fig:gainloss}) in addition to the splitting processes that are included at the present level. The splitting process is of the form $\omega'\rightarrow (\omega,\omega'-\omega)$, and should be associated with the combination of statistical factors $f(\omega') [1+f(\omega)][1+f(\omega'-\omega)]$. To this splitting corresponds a merging reverse process, where $(\omega,\omega'-\omega)\rightarrow \omega'$, associated with the combination $f(\omega)f(\omega'-\omega)[1+f(\omega')]$. In the first term of the equation the factor $f(\omega')$  should therefore be replaced by 
\beq
\Phi(\omega,\omega')&=&f(\omega') [1+f(\omega)][1+f(\omega'-\omega)]-f(\omega)f(\omega'-\omega)[1+f(\omega')]\nn
&=& f(\omega')+f(\omega')f(\omega)+f(\omega'-\omega) [f(\omega')-f(\omega)],
\eeq
where the minus sign corresponds to the fact that the merging process is a loss term (depletion of the state $\omega$), in contrast to the splitting process which populates $\omega$.
Similarly the loss splitting term $\omega\rightarrow (\omega',\omega-\omega')$ is accompanied by the combination $f(\omega)[1+f(\omega')][1+f(\omega-\omega')]$. And the corresponding merging is accompanied by the combination
$f(\omega')f(\omega-\omega')[1+f(\omega)]$. It is then easy to verify that this leads to the replacement of $f(\omega)$ in the second term of the equation by $\Phi(\omega',\omega)$. Finally, the complete equation reads
\beq\label{kinonetwo}
\frac{\partial f}{\partial t}=\frac{C}{\omega^3} \left\{   \int_\omega^\infty \rmd\omega' K(\omega,\omega') \Phi(\omega,\omega')- \int_0^\omega \rmd \omega' K(\omega',\omega) \Phi(\omega',\omega)  \right\}.
\eeq
At this point we need to specify the constant $C$. In the regime where Eq.~(\ref{kinonetwo}) applies, i.e., in the Bethe-Heitler regime, the collinear splitting yields a divergent contribution to $C$ which is regulated by a thermal mass $m$, with $m^2=m_D^2/2$, $m_D$ being the Debye mass, $m_D^2=2 g^2 N_c I_b$. After rescaling the time $t\to \tau =4\pi \alpha^2 N_c^2 t$ (that is without the Coulomb logarithm, as in  Eq.~(\ref{timescaling})), the complete calculation yields $C=RT^*$, with $R=1.83$ (see \cite{yacine} for more details on this calculation). The equation used in the main text follows from Eq.~(\ref{kinonetwo}) after a  shift the integration variable in the first integral ($p'\to p+k$), and a simple relabelling of the variables
\beq
 \label{kinonetwo2}
\partial_\tau f(p)= \frac{RT^*}{ p^3}  \left\{ \int_0^\infty \rmd k K(p,p+k) \Phi(p,p+k) -\int_0^p\rmd k K(k,p) \Phi(k,p)\right\}.
\eeq

Similar manipulations can be performed when multiple scattering play an important role and  the  LPM effect needs to be taken into account (which is not the case in the present work).  The equation for $D$ is analogous to Eq.~(\ref{eqforDBH})
\beq
\frac{\partial D(\omega)}{\partial t}=\int_x^1\rmd z \,\hat {\cal K}(z,\omega/z) D(\omega/z)-\int_0^1 \rmd z z\, \hat {\cal K}(z,\omega) D(\omega),
\eeq
the main difference being the reduced splitting kernel $\hat {\cal K}(z,\omega)$ which now reads
\beq
 \hat {\cal K}(z,\omega)=\bar\alpha\, \sqrt{ \frac{\hat q}{\omega} }\, \frac{(1-z+z^2)^{5/2}}{z^{3/2} (1-z)^{3/2}},\qquad \bar \alpha\equiv \frac{\alpha N_c}{\pi}.
 \eeq
 By ignoring the numerator (setting it equal to 1, as was done also in arriving at Eq.~(\ref{eqforDBH})), and  performing the same changes of variables as above, one obtains
\beq
\frac{\partial f}{\partial t}=\frac{\bar\alpha \sqrt{\hat q}}{\omega^3} \left\{   \int_\omega^\infty \rmd\omega' K(\omega,\omega') \Phi(\omega,\omega')- \int_0^\omega \rmd \omega' K(\omega',\omega) \Phi(\omega',\omega)  \right\},
\eeq
with now
\beq
K(\omega,\omega')=\frac{\omega'^{7/2}}{\omega^{1/2}(\omega'-\omega)^{3/2}}.
\eeq

\section{Numerical procedure}\label{numerics}

The numerical results presented in this paper where obtained for the initial density of gluons given in Eq.~(\ref{eq_glasma_f}), with 
$f_0=1$ and $Q_s=1$. 

In the numerical calculations, we use a logarithmic grid  in order to have a larger density of points in the infrared where the distribution diverges as a power law, namely, 
\beq
\ln p(i) = - \ln (p_\text{max}/p_\text{min}) \frac{n-i}{n-1} +\ln p_\text{max},
\eeq
where $i = 1,n$, $n=2000$, $p_\text{max} \equiv p(n)=10$ and $p_\text{min}\equiv p(1)=10^{-4}-10^{-3}$. 

The distribution at $\tau+\Delta \tau$  is computed as follows
\beq
f(\tau+\Delta \tau) = f(\tau) + C[f]\,  \Delta \tau,
\eeq 
with $\Delta \tau \simeq 10^{-3}$. In the case of the inelastic collision integral we use the Runge-Kutta (RK4) method to integrate over $p$ in the collision integral $C[f]$. The apparent singularity in the Kernel when $p=p'$ is regularized by requiring that $p$ never touches $p'$ in the discrete sum over $i$. Concerning the elastic part $f_\text{el}(\tau)$ we use to the Backward-Time-Centered-Space Method ``BTCS" in order to stabilize the diffusion term.

\section{Drag alone and the Burgers equation}\label{sec:burgers}

The effect of the drag current is best illustrated by ignoring the diffusion current in the kinetic equation (\ref{FPeqntau}), which reduces then to 
\beq\label{drag-eq}
\frac{\del f(p)}{\del \tau} = \frac{I_b}{p^2} \frac{\del }{\del p} \left[p^2 f(1+f) \right]\simeq \frac{I_b}{p^2} \frac{\del }{\del p} \left[p^2 f^2(p) \right],
\eeq
where in the last equality we have assumed that $f\gg 1$. In the present discussion we treat $I_b$ as a constant. 
Making the following change of variables (and expressing momenta in units of  $Q_s$ and time $\tau$ in units of $Q_s^{-1}$)
\beq
f(p)=\frac{f_0}{p} n(x),\quad t = 4 f_0 I_b \tau, \quad x =p^2,
\eeq

we transform Eq.~(\ref{drag-eq}) into
\beq\label{burgers}
\frac{\del n(x)}{\del t} = \frac{1}{2} \frac{\del }{\del x}  n^2(x),
\eeq
which is the inviscid Burgers equation in one dimension. The initial condition  reads: $n_0(x) =  \sqrt{x} \theta(1-x)$.  Note that Eq.~(\ref{drag-eq}) conserves particle number (but not energy). In fact, the Burgers equation can be written as a continuity equation, $\del n/\del t+{\rm div}j=0$,  by defining a current $j(x)=-n^2/2$.

The Burgers Eq.~(\ref{burgers}) can be solved by  the method of characteristics. This yields $n(x,t)$ as an implicit equation
\beq\label{burgers-implicit}
n (x,t)= n_0( n(x,t) t + x ).
\eeq
Given the form of the initial condition, $n_0(x)\sim \theta(1-x)$, this equation provides the solution of the Burgers equation in the region $0<x<x_\ast$, where $x_\ast(t)$  is given by 
$
n(x_\ast,t) t+x_\ast = 1.
$
In this region, Eq.~\ref{burgers-implicit} yields $n = \sqrt{n t + x} $, which for $n>0$ is solved by
$
n(x,t) = \frac{1}{2} \left[ t + \sqrt{t^2 + 4 x}\right].
$

In the original variables, the distribution function at small time and small $p$ then reads
\beq
f(p,t)=\frac{f_0}{2p}  \left[ t + \sqrt{t^2 + 4 p^2}\right].
\eeq
At $t=0$, $f(p)=f_0$, as expected. However, for any finite time, however small,  $f(p)$ is singular at $p=0$, with a behavior in $1/p$ allowing for a  flux of particles through $p=0$. Thus, in this case, condensation sets in immediately, in contrast to what happens in the Fokker-Planck equation (see the discussion in Sect.~\ref{sec:elastic}). This is of course related to the absence of a non trivial fixed point of the Burgers equation.  This equation was  obtained here by dropping the diffusion contribution to the Fokker-Planck equation, thereby eliminating the thermal fixed point (whose existence rests on the competition between drag and diffusion). Related to this, there is no notion of overoccupation for the Burgers equation, and condensation can indeed occur instantly.

\end{document}